\documentclass{aa}  
\usepackage{graphicx}
\usepackage{txfonts}
\def\kms{km~s$^{-1}$}
\def\comain{$^{12}$CO~(1--0)}
\def\coisot{$^{13}$CO~(1--0)}
\newcommand{\tsys}{$T_{\rm sys}$}
\newcommand{\um}{$\mu$m}

\newcommand{\cmdue}{cm$^{-2}$}

\newcommand{\msun}{M$_{\odot}$}

\def\her{{\it Herschel}}

\begin{document}

   \title{The Forgotten Quadrant Survey\footnote{Table 1 is only available in electronic form at the CDS via http://cdsweb.u-strasbg.fr/cgi-bin/qcat?J/A+A/}}

   \subtitle{$^{12}$CO and $^{13}$CO (1--0) survey of the Galactic plane in the range 220\degr$<l<$240\degr\, -2\fdg5$<b<$0\degr}
\titlerunning{The Forgotten Quadrant Survey}

   \author{
  M. Benedettini\inst{1}, S. Molinari\inst{1}, A. Baldeschi\inst{2,1}, M. T. Beltr\'{a}n\inst{3}, J. Brand\inst{4}, R. Cesaroni\inst{3}, D. Elia\inst{1}, F. Fontani\inst{3}, M. Merello\inst{5,1}, L. Olmi\inst{3}, S. Pezzuto\inst{1}, K. L. J. Rygl\inst{4}, E. Schisano\inst{1}, L. Testi\inst{3,6}, A. Traficante\inst{1}
          }

   \institute{
INAF -- Istituto di Astrofisica e Planetologia Spaziali, via Fosso del Cavaliere 100, 00133 Roma, Italy \email{milena.benedettini@inaf.it}
\and
Center for Interdisciplinary Exploration and Research in Astrophysics and Department of Physics and Astronomy, Northwestern University, Evanston, IL 60208
\and
INAF -- Osservatorio Astrofisico di Arcetri, Largo E. Fermi 5, 50125, Firenze, Italy 
\and
INAF -- Istituto di Radioastronomia \& Italian ALMA Regional Centre, via P. Gobetti 101, 40129, Bologna
\and
Universidade de S\~ ao Paulo, IAG Rua do Mat\~ ao, 1226, Cidade Universit\'aria, 05508-090, S\~ ao Paulo, Brazil 
\and
ESO/European Southern Observatory, Karl-Schwarzschild-Strasse 2, 85748, Garching bei M\"{u}nchen, Germany
             }
\authorrunning{Benedettini et al.}

   \date{Received ; accepted}

 
  \abstract
   {} 
   {We present the Forgotten Quadrant Survey (FQS), an ESO large project that used the 12m antenna of the Arizona Radio Observatory to map the Galactic plane in the range 220\degr$<l<$240\degr\, and -2\fdg5$<b<$0\degr, both in \comain\, and \coisot, at a spectral resolution of 0.65 \kms\, and 0.26 \kms. }
   {We used the (1--0) transition of carbon monoxide to trace the molecular component of the interstellar medium. Our data set allows us to easily identify how the molecular dense gas is organised at different spatial scales: from the giant clouds with their denser filamentary networks, down to the clumps and cores that host the new-born stars and to obtain reliable estimates of their key physical parameters such as size and mass.}
   {We present the first release of the data of the FQS survey and discuss their quality. Spectra with 0.65 \kms\, velocity channels have noise ranging from 0.8 K to 1.3 K for \comain\, and from 0.3 K to 0.6 K for \coisot. In this first paper, we used the \comain\, spectral cubes to produce a catalogue of 263 molecular clouds. The clouds are grouped in three main structures corresponding to the Local, Perseus, and Outer arms up to a distance of $\sim$8.6 kpc from the Sun. This is the first self-consistent statistical catalogue of molecular clouds of the outer Galaxy obtained with a subarcminute spatial resolution and therefore able to detect not only the classical giant molecular clouds, but also the small clouds and to resolve the cloud structure at the sub-parsec scale up to a distance of a few kiloparsec. We found two classes of objects: structures with sizes above a few parsecs that are typical molecular clouds and may be self-gravitating, and subparsec structures that cannot be in gravitational equilibrium and are likely transient or confined by external pressure. 
   We used the ratio between the \her\, H$_2$ column density and the integrated intensity of the CO lines to calculate the CO conversion factor and we found mean values of (3.3$\pm$1.4)$\times$10$^{20}$ \cmdue(K \kms)$^{-1}$ and (1.2$\pm$0.4)$\times$10$^{21}$ \cmdue(K \kms)$^{-1}$, for \comain\, and \coisot, respectively.}
   {The FQS contributes to the general effort in producing a new generation of high-quality spectroscopic data for the Galactic plane in the less-studied third Galactic quadrant toward the outer Galaxy. The FQS has produced a data set of great legacy value, largely improving the data quality both in terms of sensitivity and spatial resolution over previous data sets.}

   \keywords{ISM: clouds -- ISM: structure -- ISM: kinematics and dynamics}

   \maketitle
%

\section{Introduction}

To reconstruct the distribution of matter in our Galaxy and to understand how the diffuse interstellar medium (ISM) is structured are some of the main objectives in modern astronomy. Large surveys at all wavelengths have been carried out to map the diffuse gas content of our Galaxy as well as the densest condensations from which new stars will be born. In particular, over recent decades many projects have surveyed the Milky Way plane at several millimetre and submillimetre wavelengths with the aim of deriving the properties of the ISM through molecular spectroscopy and thermal continuum emission.

Surveys of the molecular component of the ISM play an essential role in our understanding of where and how star formation takes place in our Galaxy. The lower-frequency rotational transitions of CO are readily observed even in quite tenuous molecular gas, and the lowest of these, the (1--0) line at 115 GHz, is the preferred tracer of the cold molecular component of the ISM. The first CO surveys were carried out in the seventies, covering different parts of the Galactic plane. The first complete survey of the Galactic plane in CO (1--0) was compiled by \citet{dame1987} collecting previous partial surveys. Another extended survey covering the inner Galaxy and part of the third quadrant is the NANTEN survey \citep{mizuno2004}. 

More recently, infrared and submillimetre Galactic plane surveys were carried out both from space, such as the Galactic plane Infrared Mapping Survey Extraordinaire (GLIMPSE, \citealt{benjamin2003}), the MIPS Galactic plane survey (MIPSGAL, \citealt{carey2009}), the \her\, InfraRed Galactic plane survey (Hi-GAL, \citealt{molinari2010}), and the all-sky survey from WISE \citep{wright2010}, and from ground, such as the Bolocam Galactic Plane Survey (BGPS, \citealt{aguirre2011}), the APEX Telescope Large Area Survey of the Galaxy (ATLASGAL, \citealt{schuller2009}), and the JCMT Plane Survey (JPS, \citealt{moore2015}). With their good sensitivity and spatial resolution these surveys started to provide a picture of the distribution of dust at all scales: from diffuse clouds and filamentary structures down to pre-stellar clumps and young stellar objects (YSOs). These  multi-wavelength continuum data are ideal to derive the dust temperature of the sources, but in order to obtain the other relevant physical properties, such as the  mass, and establish whether they are in gravitational equilibrium, a good estimate of the source distance is required. The simplest way to measure distances in our Galaxy is to measure the $\varv_{\rm lsr}$ of gas from spectroscopic observations and then derive the distance by assuming a Galactic rotation model. Furthermore, the identification of possible multiple molecular structures of the ISM at different distances along the same line of sight and the characteristics of their dynamical state can only be derived from spectroscopic observations.

The first generation of CO surveys are at very poor spatial resolution (a few arcminutes) and coarse spatial sampling, making it very difficult to use them in conjunction with the most recent continuum surveys with spatial resolution from a few to tens of arcseconds. Therefore, the production of a new generation of unbiased surveys of the molecular gas with resolution and sensitivity comparable to those of the continuum surveys has become mandatory. Several projects have been started in recent years to map large portions of the Milky Way in CO and its isotopologues, at high spatial (sub-arcmin) and spectral (a few tenths of a \kms) resolution. The largest spectroscopic surveys are: the Galactic Ring Survey (GRS, \citealt{jackson2006}), the Mopra Southern Galactic Plane CO survey \citep{burton2013}, the Exeter-FCRAO CO  Galactic Plane Survey \citep{mottram2010}, the Three-mm Ultimate Mopra Milky Way Survey \citep{barnes2015}, the Structure, Excitation, and Dynamics of the Inner Galactic Inter-Stellar Medium (SEDIGISM) survey \citep{schuller2017}, The Milky Way Image Scroll Painting \citep{jiang2013}, and the FOREST Unbiased Galactic plane Imaging survey with the Nobeyama 45-m telescope (FUGIN, \citealt{umemoto2017}). 
The third Galactic quadrant is the least observed, with several pieces of the Galactic plane not yet covered by ongoing projects \citep{heyer2015}. The Forgotten Quadrant Survey (FQS) aims at covering this gap by mapping a  strip in the range 220\degr $ < l <$ 240\degr\, and  -2\fdg5$<b<$0\degr, centred around a midplane that follows the Galactic warp, both in $^{12}$CO and $^{13}$CO (1--0) transitions. The longitude range covers the outer Galaxy, crossing the Local arm, the Perseus arm, and the Outer arm, a region still densely populated with a variety of star-forming regions and complex networks of filamentary structures \citep{elia2013}. 

The FQS is an ESO project awarded 700 hrs of observing time at the Arizona Radio Observatory (ARO) 12m antenna. The primary aim of the project is the characterisation of the molecular gas in this poorly studied portion of the third quadrant of the Milky Way. To achieve this goal we will produce catalogues of the main structures identified in the FQS spectroscopic data at different spatial scales, from large molecular clouds to dense cores and we will determine their distance and derive their main physical properties, such as size and mass. 
In particular, the main scientific goals of the survey are the following.

{\it i.} To produce a catalogue of molecular clouds (MCs) in order to get a better understanding of the large-scale structure in this portion of the outer Galaxy, with improved spatial resolution.
The FQS data allow the identification of MCs, disentangling clouds at different distances along the same line of sight. The simultaneous observation of the $^{12}$CO and $^{13}$CO isotopologues allows the computation of a reliable estimate of the CO column density and mass for each cloud.
 
{\it ii.} To produce a catalogue of filamentary structures and to investigate their formation mechanism. \her\, photometric surveys have shown that  a large part of the ISM is structured in filaments and a catalogue of filament candidates of the Galactic plane from the \her\, data was published (Schisano et al., submitted).
The FQS spectroscopic data allow us to verify the kinematic coherency of the filament candidates identified in the Hi-GAL photometric maps. For these filaments the velocity field can be measured, allowing the study of their internal dynamics and feeding mechanisms.

{\it iii.} To produce a catalogue of dense cores and build the core mass function (CMF). The spectral resolution of the FQS data offers the possibility to measure the non-thermal contribution to the line broadening and thus to measure the virial mass of dense cores and clumps. The comparison of the virial mass with the core or clump mass derived from submillimetre continuum surveys (e.g. the Hi-GAL catalogue) determines whether they are gravitationally bound or unbound. This is of utmost importance for distinguishing structures that may eventually form a star or a multiple stellar system from those that will disperse, and thus for constructing a correct CMF. 

{\it iv.} To derive distance estimates for the dense compact sources detected in the continuum far-infrared maps. The correct exploitation of catalogues of compact and potentially star-forming clumps or cores produced by IR and submillimetre continuum surveys (e.g. the Hi-GAL catalogue) is based on the possibility to derive a reliable mass estimate, which in turn depends on a correct distance measurement. The FQS data provide the $\varv_{\rm lsr}$ of the emitting gas from which its distance can be derived by assuming a Galactic rotation model. The comparable spatial resolution between FQS and Hi-GAL data allows one to easily associate this distance to the Hi-GAL sources.

All these scientific topics will be investigated in several dedicated papers. In this first paper we present the FQS project with its data products and we address the first item of our listed specific goals. Indeed, we present the first self-consistent catalogue of molecular clouds of the outer Galaxy extracted from fully sampled \comain\, maps at subarcminute spatial resolution. The structure of the paper is as follows. Section 2 describes the observational setup and the data-acquisition strategy. Section 3 presents our data-reduction pipeline with the final products and the quality assessment. In Sect. 4 we describe the \comain\, and \coisot\, line emission maps. The MCs catalogue is presented in Sect. 5 with the extraction algorithm and the derived physical properties. In Sect. 6  we focus on the nature of the subparsec-scale structures and in Sect. 7 we show how the MCs are distributed in the plane of the Milky Way. Section 8 contains a discussion of the CO-to-H$_2$ conversion factor and conclusions are drawn in Sect. 9.


  \begin{figure*}
   \centering
      \includegraphics[width=0.95\textwidth]{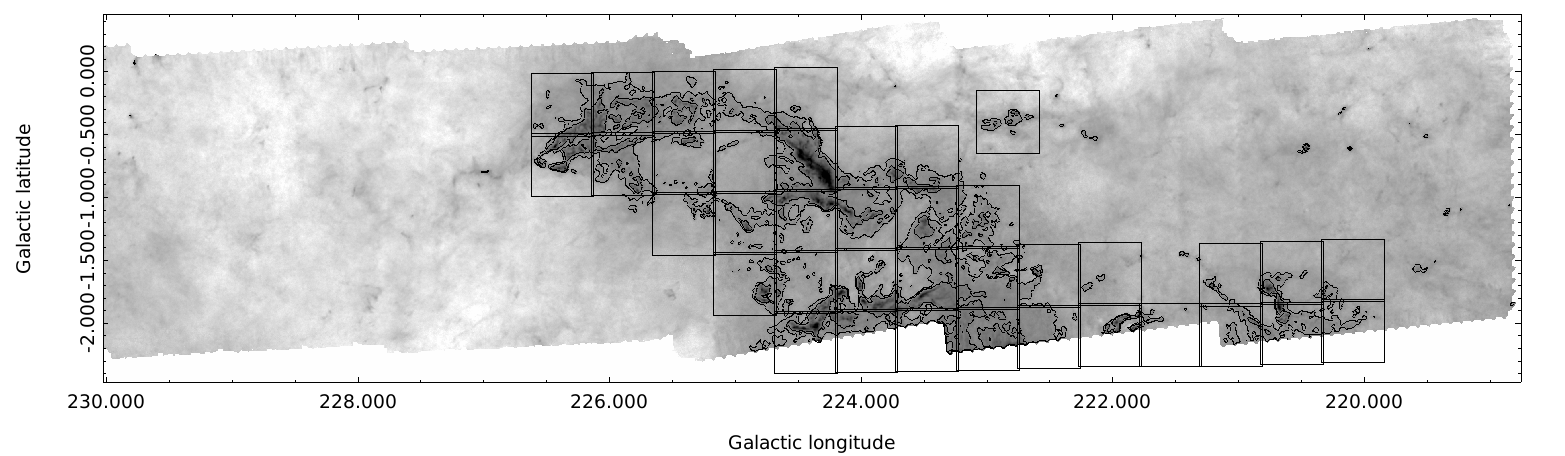}          
      \includegraphics[width=0.95\textwidth]{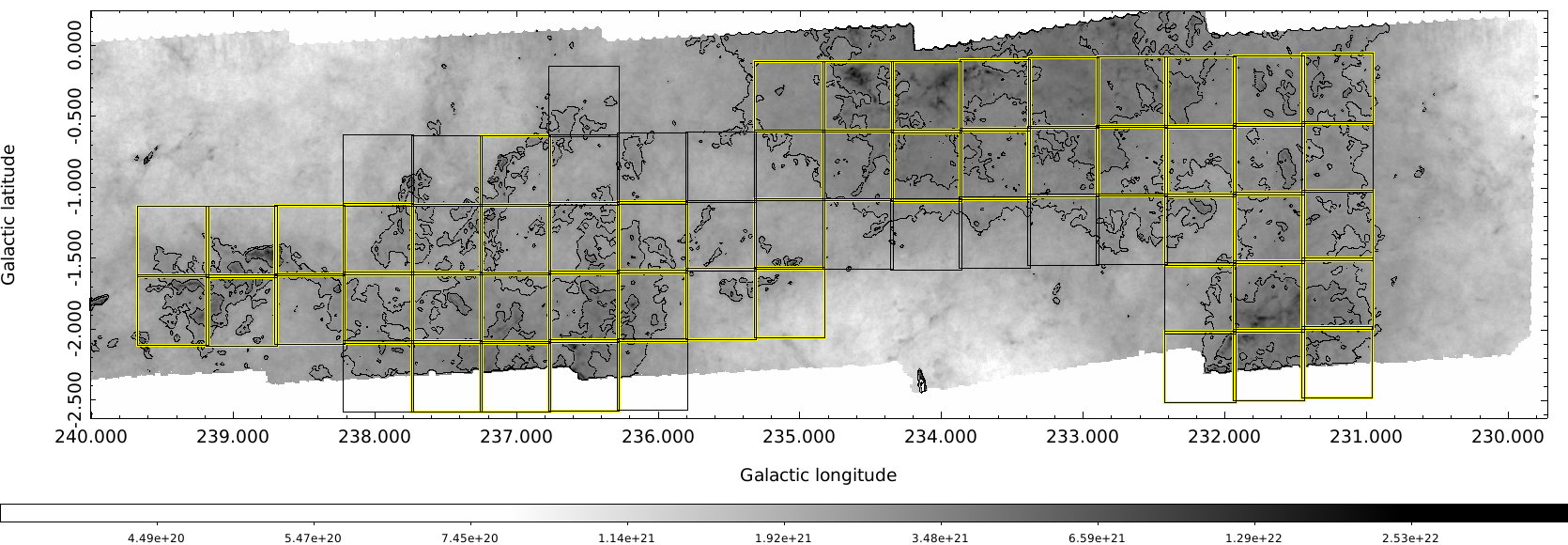}
      \caption{Coverage of the FQS survey with the observed 30\arcmin$\times$30\arcmin\, tiles (squares) overplotted on the H$_2$ column density maps derived from \her\, data. Top panel: Range of Galactic longitude 219\degr--230\degr, the coverage of \comain\, and \coisot\, data (black squares) is the same. Bottom panel: Range 230\degr--240\degr, where the coverage of the \coisot\, data (yellow squares) is smaller than the coverage of the \comain\, data (composed by yellow squares plus black squares). The intensity bar of the H$_2$ column density at the bottom is in units of \cmdue. The two contours correspond to $N(\rm H_2)$ = 2.8 $\times$10$^{21}$ \cmdue\, and 3.8 $\times$10$^{21}$ \cmdue, equivalent to a visual extinction level of 3 and 4 mag, respectively. The sensitivity level of our observations allows detection of signal only from regions inside the contour at about 3 mag for \comain\, and at about 4 mag for \coisot. }
      \label{fig:coverage}
   \end{figure*}

\section{Observations}

\subsection{Observational setup}

We used the ARO AEM ALMA prototype 12m antenna with the ALMA Type Band-3 Receiver to map the Milky Way plane in the range 220\degr$<l<$ 240\degr\, and -2\fdg5$<b<$0\degr, following the Galactic warp, both in \comain\, and \coisot. Observations started in November 2016 and ended in June 2018.

The survey area was divided into partially overlapping tiles of 30\arcmin$\times$30\arcmin\, size, with sides aligned along the Galactic longitude and latitude (see Fig. \ref{fig:coverage}). Each tile was observed twice in the On-The-Fly observing mode in mutually orthogonal scan directions, one along $l$ and one along $b$. The data dumping time is 0.1 s. The distance of the scanning rows is 18\arcsec\, and the scanning speed is 75$^{\prime\prime}$/s, resulting in a better-than Nyquist spatial sampling of the telescope beam, that is 55\arcsec\, at 115 GHz. Several positions for the OFF measurement were used, so that for each tile the distance between the centre of the tile and the OFF position was smaller than 2\degr. All the OFF positions were checked to be free of \comain\  emission.

The receiver was tuned to 115.271 GHz and 110.201 GHz for \comain\, and \coisot, respectively. The back-end was composed of a 256-channel filter bank at 250 kHz spectral resolution (hereafter FB250), corresponding to a total velocity coverage of 166 \kms\, and a velocity resolution of 0.65 \kms, in parallel with a second 256-channel filter bank at 100 kHz spectral resolution (hereafter FB100), corresponding to a total velocity coverage of 66 \kms\, and a velocity resolution of 0.26 \kms.
Antenna temperature is converted to main beam temperature by dividing by the beam efficiency of 0.80.

\subsection{Calibration accuracy and stability}

In order to evaluate the temperature calibration accuracy and to monitor the stability of the system over the long period of data acquisition, at the beginning of each observing day we acquired the spectrum of a reference source. We observed IRAS 07046-1115 at RA(J2000) = 7$^h$ 07$^m$ 01.6$^s$ and DEC(J2000) = -11\degr\, 20\arcmin\, 17\arcsec. This source was previously observed in \comain\, at SEST by \citet{wouterloot1989}.
The total integrated intensity of the \comain\, line towards IRAS 07046-1115 measured during the observation campaign with FB250 has a mean value of $\int T^*_{\rm A} {\rm d}\varv$ = ($71.5 \pm 1.9$)~K~\kms; for the spectra of FB100 the mean value is ($70.8 \pm 1.9$)~K~\kms. The two measurements are in agreement within the uncertainties. We do not observe any systematic trend in time of the integrated intensity measurements and the deviations of the measured integrated intensity with respect to its mean value are within $\sim$5$\%$, for both filter banks, smaller than the absolute temperature calibration accuracy of the ARO 12m data which is $\sim 10\%$.

\begin{figure}
   \centering
    \includegraphics[width=0.45\textwidth]{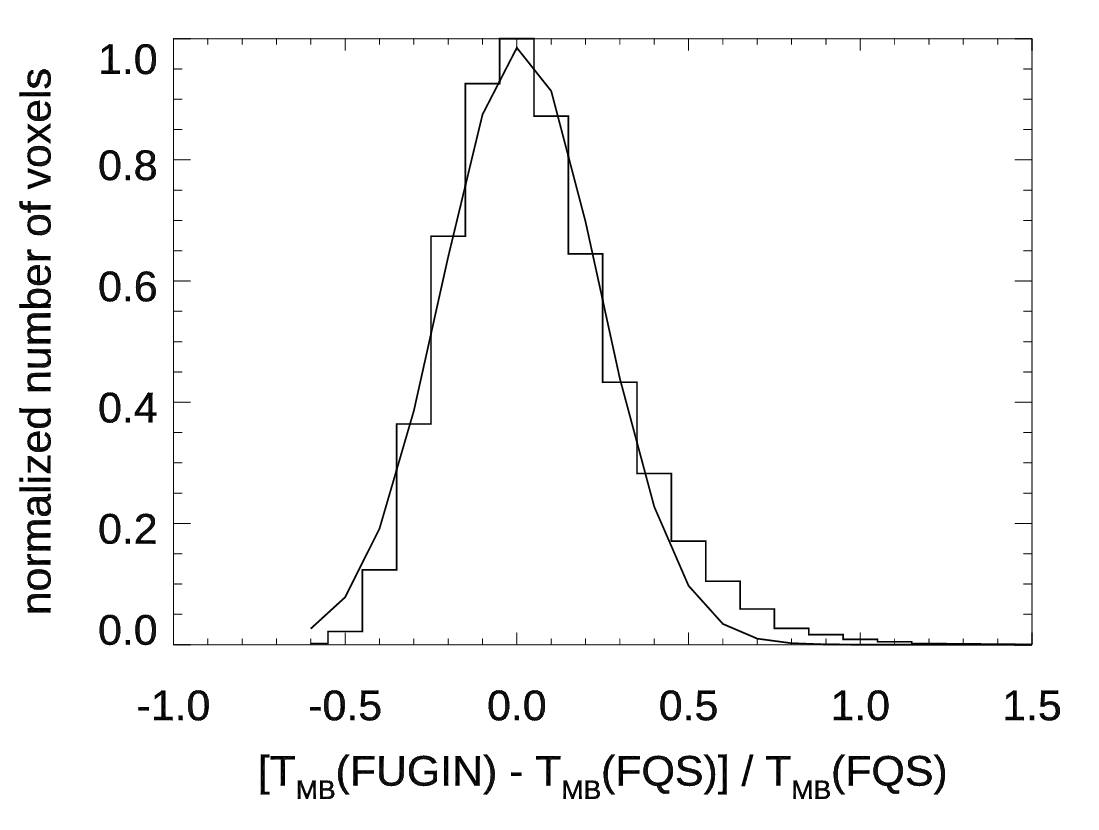}
    \caption{Histogram of the normalised relative difference between the main beam temperature of the \comain\, line measured in the FQS and in the FUGIN survey. The data are resampled to a common spatial resolution (55\arcsec), pixel size (1717\farcs3), and velocity channel (1 \kms). All voxels ($l,b,\varv$) with emission above 5$\sigma$ for both sets of data are considered.The continuum line shows the best-fit Gaussian distribution with peak at 0.01 and standard deviation of 0.23. }
   \label{fig:calib}
   \end{figure}

In order to crosscheck the absolute temperature calibration of the ARO 12m data, we compared the ARO measurement with the same measurement of the \comain\, line taken with other telescopes. Firstly, we used the observation made with SEST towards IRAS 07046-1115. Converting the antenna temperature to main beam temperature by assuming a beam efficiency of 0.80 results in an integrated intensity measured at ARO 12m of $\int T_{\rm MB} {\rm d}\varv$ = ($89 \pm 2.4$)~K\kms. The SEST data give $\int T_{\rm MB} {\rm d}\varv$ = 93.6 K \kms, a slightly higher value that could, at least partially, be ascribed to the different beam of the two telescopes, namely  55\arcsec\, for ARO 12m and 43\arcsec\, for SEST and that, in any case, is within the absolute calibration accuracy. An additional comparison was done with the FUGIN survey \citep{umemoto2017}, that partially overlaps with FQS in the range 220$\degr < l <$ 236\degr, -1\degr$< b <$ 0\degr. To this aim, we considered the common area of one square degree at 225$\degr < l <$ 226\degr\, and -1\degr$< b <$ 0\degr. The Nobeyama 45m Fugin \comain\, data were convolved to the beam size of the ARO 12m, rebinned to a common velocity channel of 1 \kms\, and regridded to the FQS spatial grid. Then, for each voxel ($l,b,\varv$) with $T_{\rm MB}$ of the \comain\, line above 5$\sigma$ for both sets of data, we calculated the relative main beam temperature difference, $[T_{\rm MB}(\rm FUGIN) - T_{\rm MB}(\rm FQS)] / T_{\rm MB}(\rm FQS)$, whose histogram is shown in Fig. \ref{fig:calib}. The distribution is well fitted by a Gaussian function with peak at 0.01 and standard deviation of 0.23. The tail at the highest relative difference, namely the voxels with relative difference $\geq$ 0.6, is a statistical effect associated to voxels with the lowest signal-to-noise ratios and they are only 4\% of the total. In conclusion, there is no systematic difference between the two sets of data and in about 68\% of the voxles the FQS and FUGIN data are consistent within 23\%, in agreement with the intensity variations measured towards the reference source by FQS (5\%) and FUGIN (20\%, \citealt{umemoto2017}).

\subsection{Mapped area}

The selected portion of the Galactic plane, namely 220\degr $< l <$ 240\degr\, and -2\fdg5$<b<$0\degr, was not fully covered by our ARO observations. In fact, we used the cold dust emission detected in the \her\, maps as a proxy of the CO emission. We compared the \comain\ integrated line maps in an initial set of tiles with the H$_2$ column density maps derived from \her\, data (Schisano et al., submitted) and found that, given the sensitivity level of our observations, the \comain\ line was detected only in regions with H$_2$ column densities above $\sim$ 2.8 $\times$10$^{21}$ \cmdue, equivalent to a visual extinction level of $A_{\rm V}\geq$ 3 mag, assuming the conversion factor $N$(H$_2$)/$A_{\rm V}$ = 9.4$\times$10$^{20}$ \cmdue mag$^{-1}$ \citep{bohlin78}. For the \coisot\, line a similar level was found of $\sim$ 4 mag. We therefore used the 3 mag and 4 mag contour level of the \her\, H$_2$ column density map as a criterion for identifying the regions to be observed in \comain\, and \coisot, respectively. In the area not covered by the ARO observations we are confident that no  significant CO emission is present above the sensitivity level of our survey. This statement has been verified a posteriori, as discussed in Sect. \ref{sect:emission_map}. In Fig. \ref{fig:coverage} we show the total coverage in \comain\, and in \coisot\, overplotted on the \her\, H$_2$ column density map.

\section{Data reduction}

\subsection{Data reduction pipeline}

Data reduction was carried out with a dedicated pipeline developed with the GILDAS package\footnote{The GILDAS software has been developed at IRAM and Observatoire de Grenoble: http://www.iram.fr/IRAMFR/GILDAS}, which requires minimal human intervention. After manually flagging spectra affected by obvious artefacts, the data were automatically processed tile by tile. For each tile, all the spectra were averaged to obtain a low-noise spectrum, which was used to identify the velocity intervals where emission is detected. The remaining channels were used to fit a first-order polynomial to the baseline, spectrum by spectrum, to obtain baseline-subtracted data.
For a given molecular line and back-end, we merged all the spectra after resampling them both in space and velocity through standard commands in GILDAS to create spectral cubes. In this way, for both \comain\, and \coisot\, data, we obtained two cubes with different spectral resolutions: one with channels of 0.3~\kms, from the FB100 back-end, and the other with 1~\kms\, channels, from the FB250 back-end, slightly degrading the native spectral resolution of the spectra in order to increase the signal-to-noise ratio. The spatial pixel size for both cubes is 17\farcs3. To produce a manageable set of maps, we divided the survey area into two fields, one covering the Galactic longitude range  220\degr$<l<$226\fdg6 and the other covering the range 231\degr$<l<$239\fdg7, and we created separate spectral cubes for the two fields. 

Calibrated single spectra and spectral cubes are publicly available in the FQS web page\footnote{http://fqs.iaps.inaf.it}. Calibrated spectra for each 30\arcmin$\times$30\arcmin\, tile are packed into eight files in the GILDAS format, distinguishing between observed line (\comain\, or \coisot), scan direction (along $l$ or $b$), and filter banks (FB250 and FB100). The FQS spectral cubes are made available to the community also in the framework of the VIALACTEA\footnote{http://vialactea.iaps.inaf.it} knowledge base \citep{molinaro2016}, accessible through the VIALACTEA visual analytics tool \citep{vitello2018}, which allows the inspection of our data in combination with other continuum and spectroscopic data of the Galactic plane from many surveys from infrared to radio frequencies.

\subsection{Data quality}

\begin{figure}
   \centering
    \includegraphics[width=0.45\textwidth]{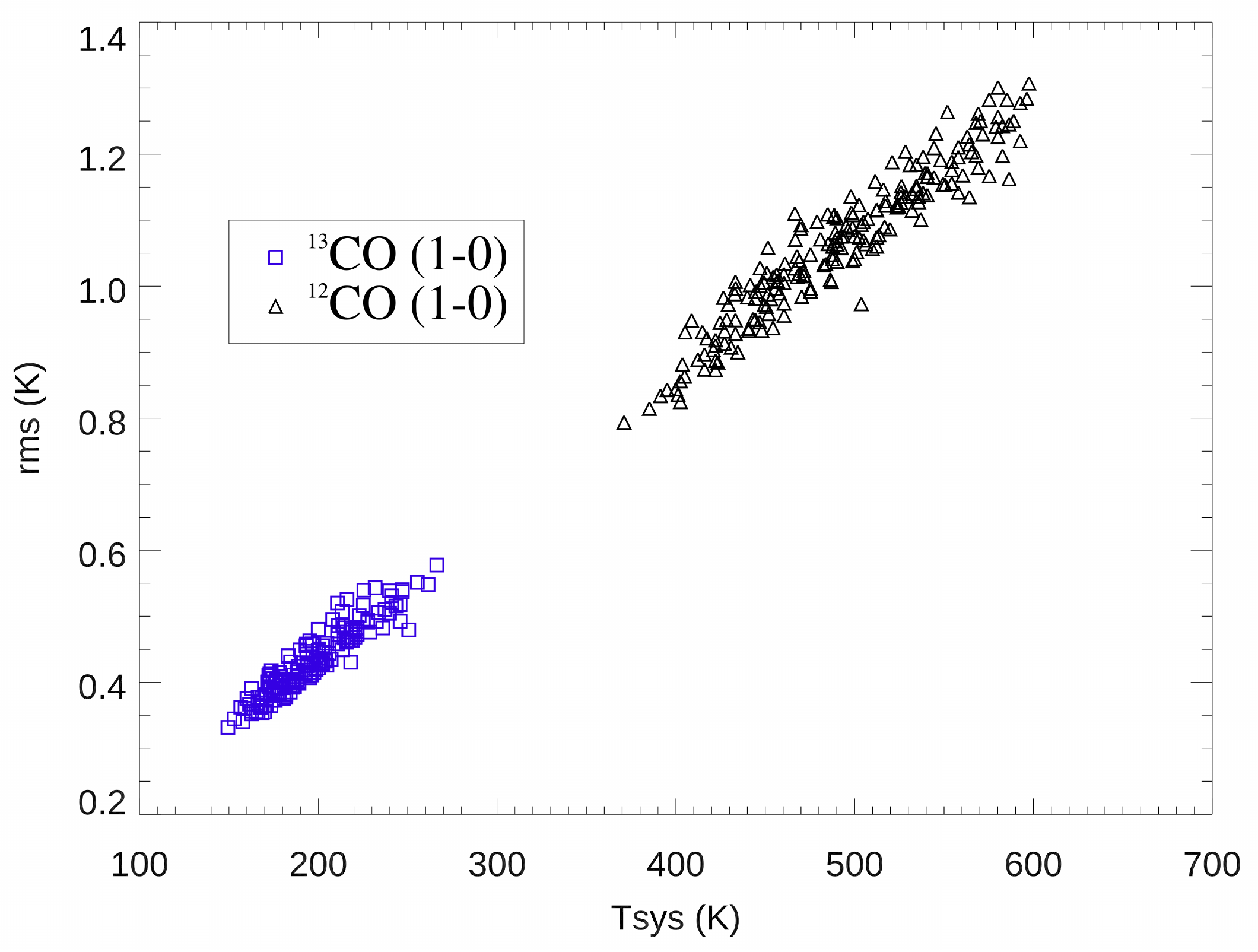}
    \caption{Median values of rms {\it vs.} \tsys\, of the single tiles observed with FB250 in each scan direction separately. Spectra are calibrated in main beam temperature and have a velocity channel of 0.65 \kms. Triangles are for \comain\, and squares for \coisot.}
   \label{fig:tsys_rms}
   \end{figure}

\begin{figure}
   \centering
    \includegraphics[width=0.45\textwidth]{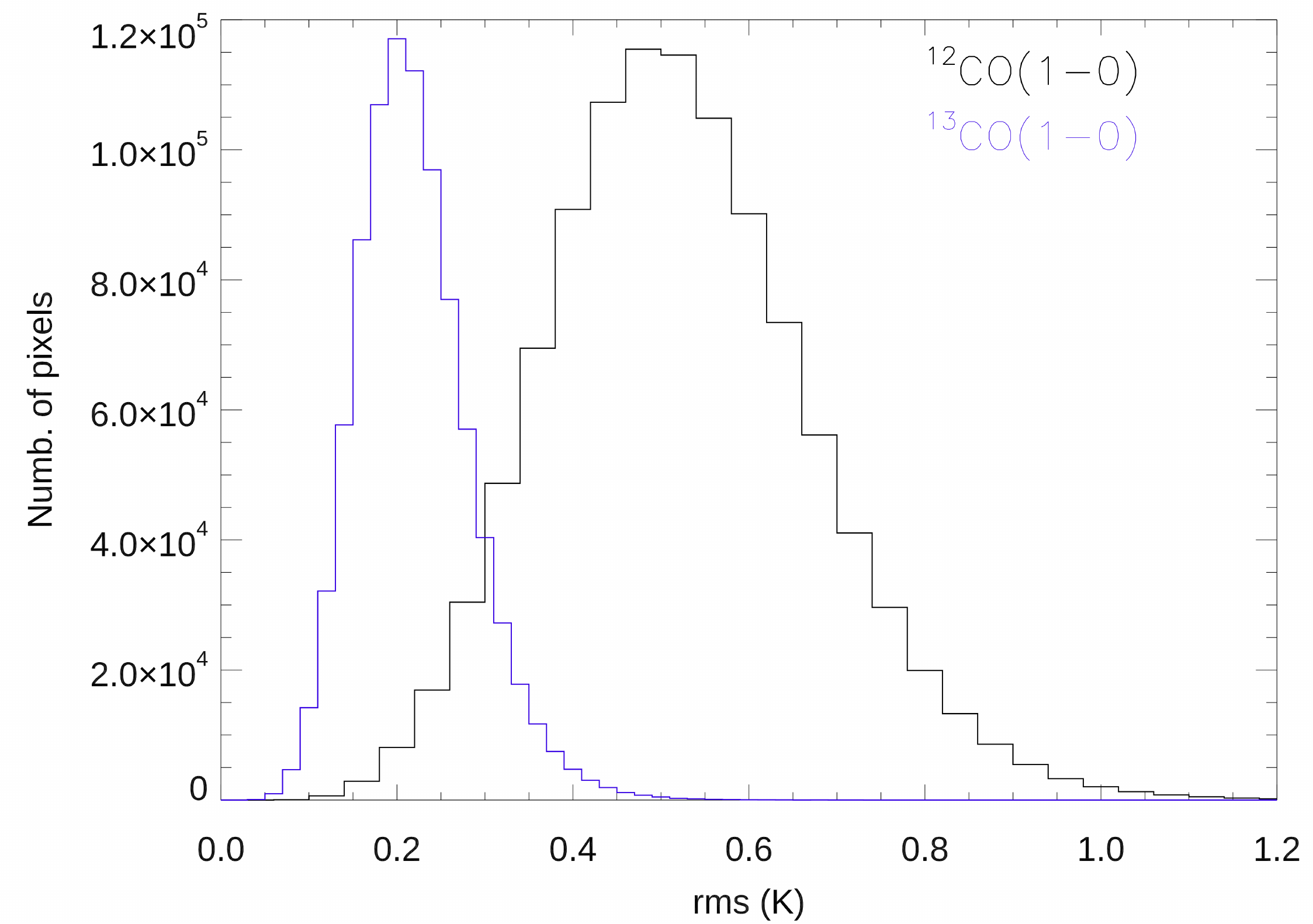}
    \caption{Histograms of rms of the spectra observed in FB250 in each spatial pixel of the final spectral cubes resampled in 1 \kms velocity channels. Blue is for \coisot\, and black for \comain.}
   \label{fig:rms_hist}
   \end{figure}

A data quality assessment was performed on the acquired data using the spectra of FB250. We spatially resampled the spectra of the 30\arcmin$\times$ 30\arcmin\, tile acquired in a single scan direction in spatial pixels of 17\farcs3, equivalent to about one-third of the beam. For each of these tiles we calculated the median value of the system temperature (\tsys) and the root mean square (rms) noise in units of main beam brightness temperature, in a reference velocity channel, at the native spectral resolution of  0.65 \kms. Maps acquired during unfavourable weather conditions were identified as outliers with respect to the average distributions. In order to guarantee an acceptable homogeneity in survey sensitivity we discarded the maps of the noisy tiles and
re-observed those tiles. In Fig. \ref{fig:tsys_rms} we show the median value of the rms, in $T_{MB}$, of the single tiles observed in one scan direction at the spectral resolution of  0.65 \kms\, plotted as a function of the median value of the corresponding \tsys\, for all of the good tiles. The rms ranges from 0.79 K to 1.31 K for \comain\, and from 0.33 K to 0.58 K for \coisot. We note that the rms of our final spectral cubes is lower since each spatial pixel is observed at least twice and the data were smoothed to a spectral resolution of 1~\kms. The histogram of the rms measured in each spatial pixel in the final cubes for both \comain\, and \coisot\, data is shown in Fig. \ref{fig:rms_hist}. The median value of the rms in the spectral cubes of the \comain\, is 0.53 K and the median rms for the cubes of the \coisot\, is 0.22 K. Comparing our sensitivity level to that of the FUGIN survey \citep{umemoto2017}, the other existing CO survey that partially overlaps with FQS, we find that our sensitivity level is about a factor of 0.4 better.

\section{$^{12}$CO and $^{13}$CO (1--0) emission maps}
\label{sect:emission_map}

  \begin{figure*}
   \centering
    \includegraphics[width=0.9\textwidth]{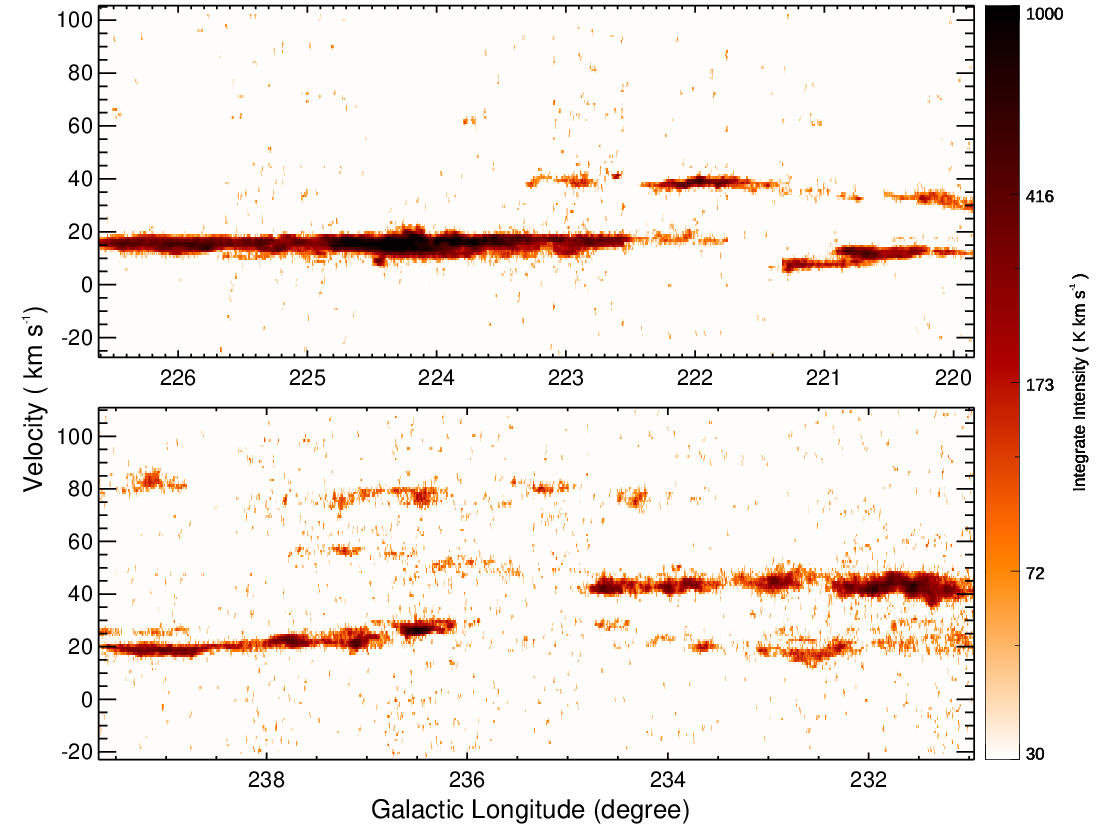}
     \caption{Galactic longitude--velocity diagrams of the \comain\, line for the two ranges of Galactic longitude 220\degr--226\fdg6 (top panel) and 231\degr--239\fdg7 (bottom panel). 
    }
   \label{fig:lv_map}
   \end{figure*}

\begin{figure*}
   \centering
   \includegraphics[width=0.83\textwidth]{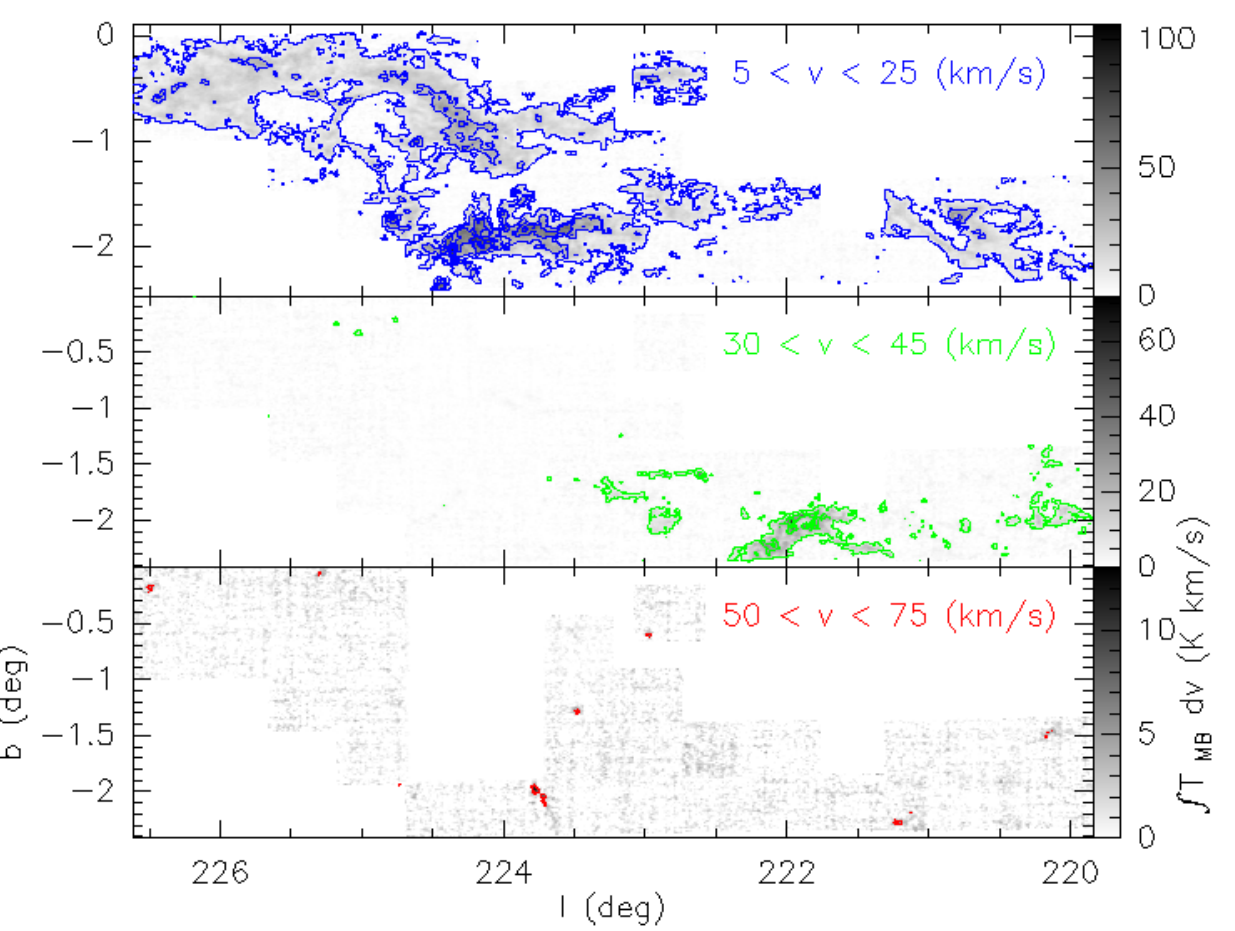}
    \caption{\comain\, intensity maps of the $l$ = 220\degr--226\fdg6 field, integrated over velocity ranges 5 -- 25 \kms\, (top panel), 30 -- 45 \kms\, (central panel), and 50 -- 75 \kms\, (bottom panel). First contour level is at 5 K \kms\, and steps are of 20 K \kms. 
    }
   \label{fig:chan_map_l224_12}
   \end{figure*}

\begin{figure*}
   \centering
    \includegraphics[width=0.83\textwidth]{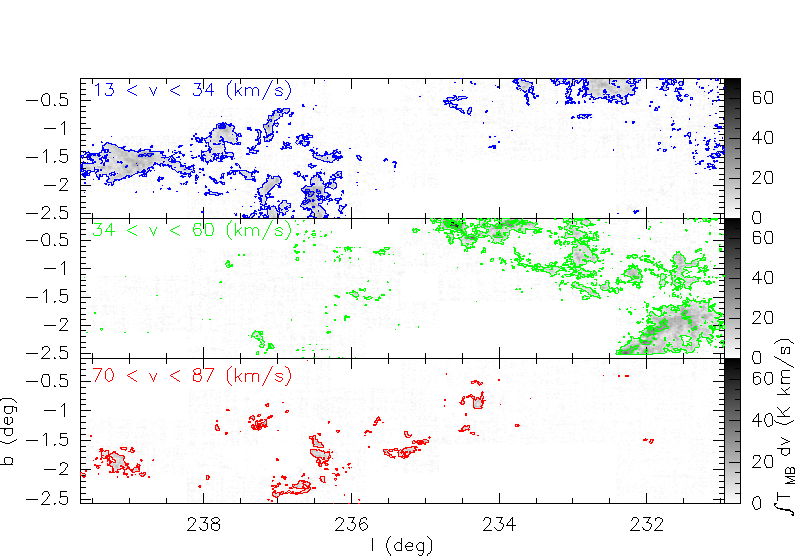}
    \caption{\comain\, intensity maps of the $l$ = 231\degr--239\fdg7 field, integrated over velocity ranges 13 -- 34 \kms\, (top panel), 34 -- 60 \kms\, (central panel), and 70 -- 87 \kms\, (bottom panel). First contour level is at 5 K \kms\, and steps are of 20 K \kms. 
    }
   \label{fig:chan_map_l235_12}
   \end{figure*}

\begin{figure*}
   \centering
    \includegraphics[width=0.83\textwidth]{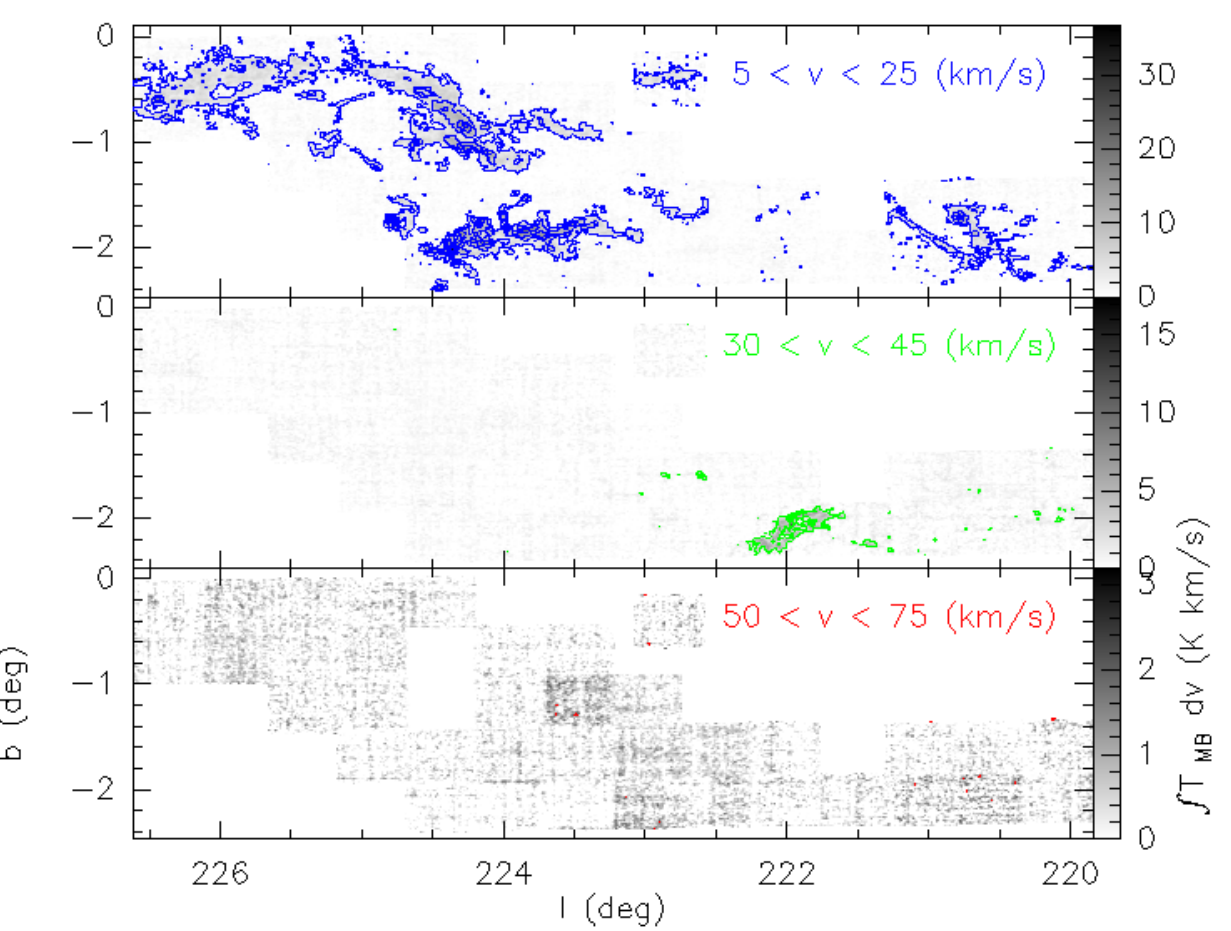}
    \caption{\coisot\, intensity maps of the $l$ = 220\degr--226\fdg6 field, integrated over velocity ranges 5 -- 25 \kms\, (top panel), 30 -- 45 \kms\, (central panel), and 50 -- 75 \kms\, (bottom panel). First contour level is at 2 K \kms\, and steps are of 5 K \kms. 
    }
   \label{fig:chan_map_l224_13}
   \end{figure*}

\begin{figure*}
   \centering
    \includegraphics[width=0.83\textwidth]{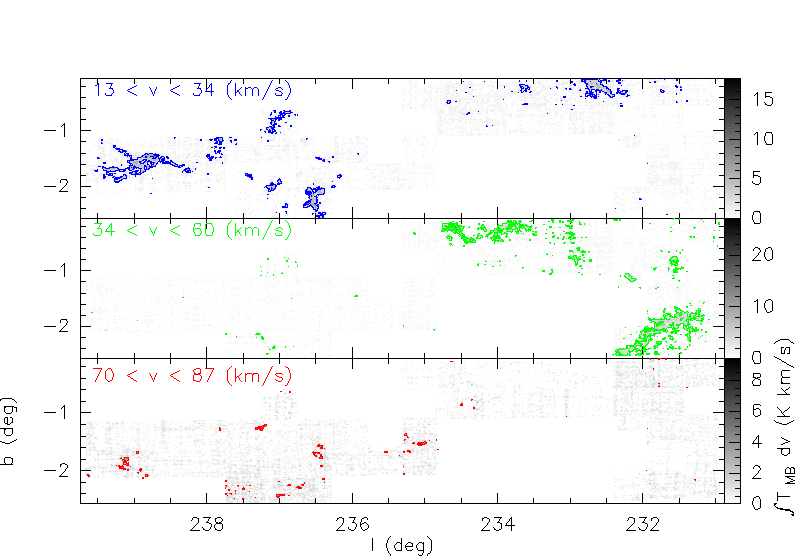}
    \caption{\coisot\, intensity maps of the $l$ = 231\degr--239\fdg7 field, integrated over velocity ranges 13 -- 34 \kms\, (top panel), 34 -- 60 \kms\, (central panel), and 70 -- 87 \kms\, (bottom panel). First contour level is at 2 K \kms\, and steps are of 5 K \kms. 
    }
   \label{fig:chan_map_l235_13}
   \end{figure*}

\begin{figure*}
   \centering
      \includegraphics[width=0.95\textwidth]{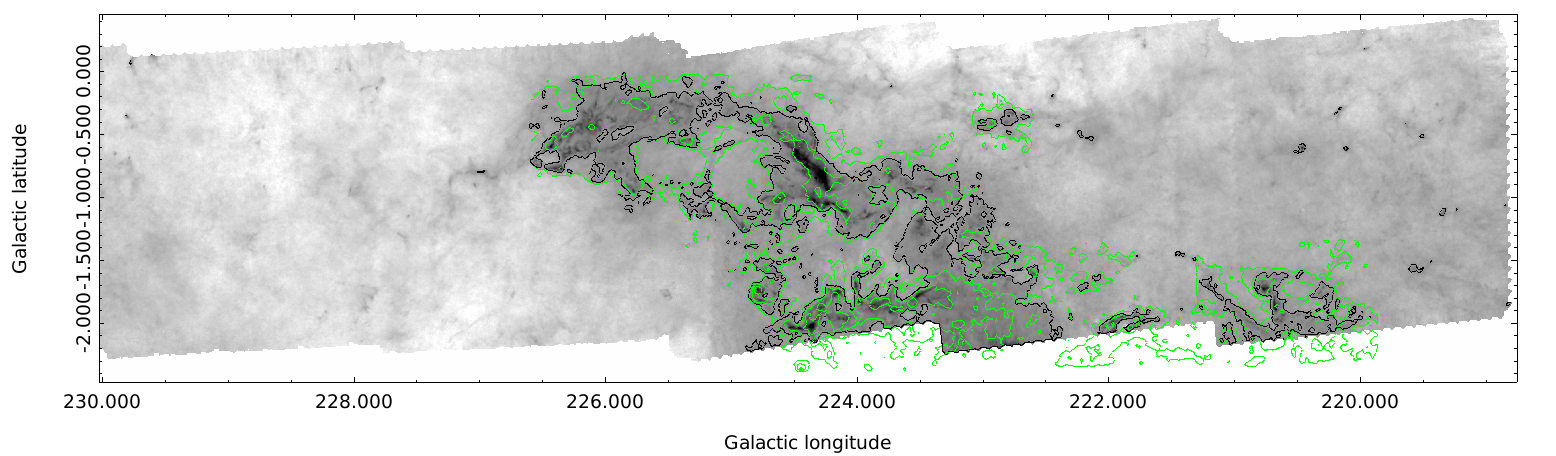}      
      \includegraphics[width=0.95\textwidth]{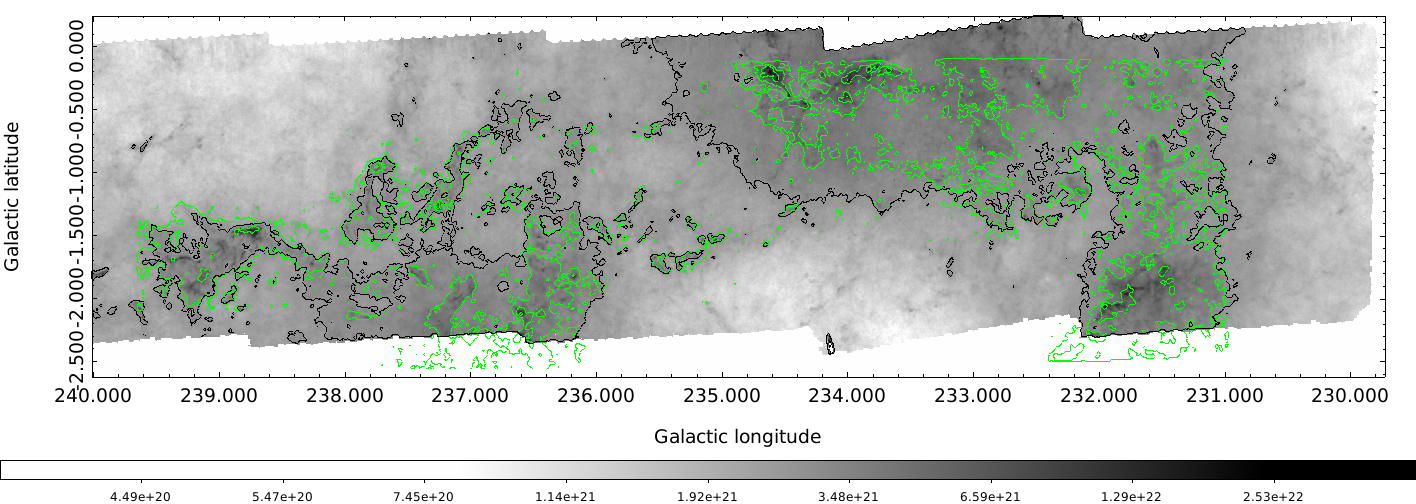}      
      \caption{H$_2$ column density maps (grey scale and black contour) derived from \her\, data with the \comain\, intensity maps overplotted as green contours. Top panel: Range of Galactic longitude 219\degr--230\degr; the \comain\, emission is integrated over the velocity range 5 -- 75 \kms; green contours are at 5, 25 and 50 K \kms. Bottom panel: Range 230\degr--240\degr; the \comain\, emission is integrated over the velocity range 13 -- 87 \kms; green contours are at 5, 25, and 50 K \kms. The intensity bar of the H$_2$ column density at the bottom is in units of \cmdue. The black contour corresponds to $N(\rm H_2)$ = 2.8 $\times$10$^{21}$ \cmdue, equivalent to a visual extinction level $A_{\rm V}$ = 3 mag. }
         \label{fig:h2_12co}
   \end{figure*}

In Fig. \ref{fig:lv_map} we present the velocity  versus Galactic longitude plot of the \comain\, line, obtained by integrating the emission over the observed range of Galactic latitude (-2\fdg5 $<b<$ 0\degr). Three main structures are clearly visible as well-separated, semi-continuous horizontal features, each of which contains molecular gas with similar velocity. These features correspond to three spiral arms of the Galaxy: the first, with velocities between $\sim$ 5 \kms\, and $\sim$ 30 \kms, corresponds to the Local arm, the second, with velocities between $\sim$ 30 \kms\, and $\sim$ 60 \kms, corresponds to the Perseus arm, and the third, with velocities between $\sim$ 60 \kms\, and $\sim$ 90 \kms, corresponds to the Outer arm (see Sect. \ref{sect:pos_gal} for further details). In Figs. \ref{fig:chan_map_l224_12} and \ref{fig:chan_map_l235_12} we plot the intensity maps of the \comain\, line, integrated over the three velocity ranges. These show the spatial distribution of the molecular gas in the observed portion of the Galactic plane. In several parts of the plane, multiple clouds are present along the same line of sight, but they are easily disentangled in the spectra because their velocities are well separated (see Fig. \ref{fig:lv_map}). The intensity maps of \coisot\, integrated over the same velocity ranges as for \comain\  are presented in Figs. \ref{fig:chan_map_l224_13} and \ref{fig:chan_map_l235_13}. Although the sensitivity level of our \coisot\, maps is about a factor of two better than that of the \comain\, maps, the \coisot\, emission traces only the central, densest part of the molecular gas, because of the higher critical density and the lower chemical abundance of $^{13}$CO.
In Fig. \ref{fig:h2_12co} we compare the \her\, H$_2$ column density map with the \comain\, integrated intensity map. The H$_2$ column density contour of 2.8$\times$10$^{21}$ \cmdue\  that we used as proxy for the \comain\, line emission matches well with the 3$\sigma$ level of the CO map, except a few regions where it encompasses a slightly more extended area with respect to that emitting in CO. Therefore, the coverage of our CO survey, although patchy, catches all the emission of the observed lines above the sensitivity level of our observations in the Galactic plane in the range 220\degr$<l<$240\degr\, and -2\fdg5$<b<$0\degr, with the very limited exception of a handful of small and isolated spots.

We note that several structures seem to extend beyond the coverage of our survey and therefore they may not be fully mapped in our survey, in particular at ($l\sim$222\fdg3, $b<$-2\fdg5); ($l\sim$224\fdg5, $b<$-2\fdg5); ($l\sim$232\degr, $b<$-2\fdg5); ($l\sim$232\fdg35, $b>$0\degr); ($l\sim$234\degr, $b>$0\degr), and ($l\sim$237\degr, $b<$-2\fdg5). 

\section{Molecular cloud catalogue}

\subsection{Cloud identifications}

To reconstruct the large-scale distribution of the molecular gas in the survey area, we produced a catalogue of molecular clouds using the Spectral Clustering for Interstellar Molecular Emission Segmentation ({\it SCIMES}) algorithm \citep{colombo2015}. The algorithm is based on a cluster analysis of dendrograms of 3D ({\it l-b-v}) data cubes that performs particularly well for the identification of large structures such as the MCs (e.g. \citealt{colombo2015,schuller2017}). {\it SCIMES} considers the dendrogram tree of the 3D structures in the data cube (using the implementation of \citealt{rosolowsky2008} for astronomical data sets) and groups different leaves together into ‘clusters’ of leaves based on one or more criteria. The algorithm requires the definition of some parameters that regulate the level of clustering. We tuned these values to maximise the connection of the identified structures in large, kinematically coherent clouds (see below for the numerical value of the parameters).

\begin{figure}
   \centering
    \includegraphics[width=9cm]{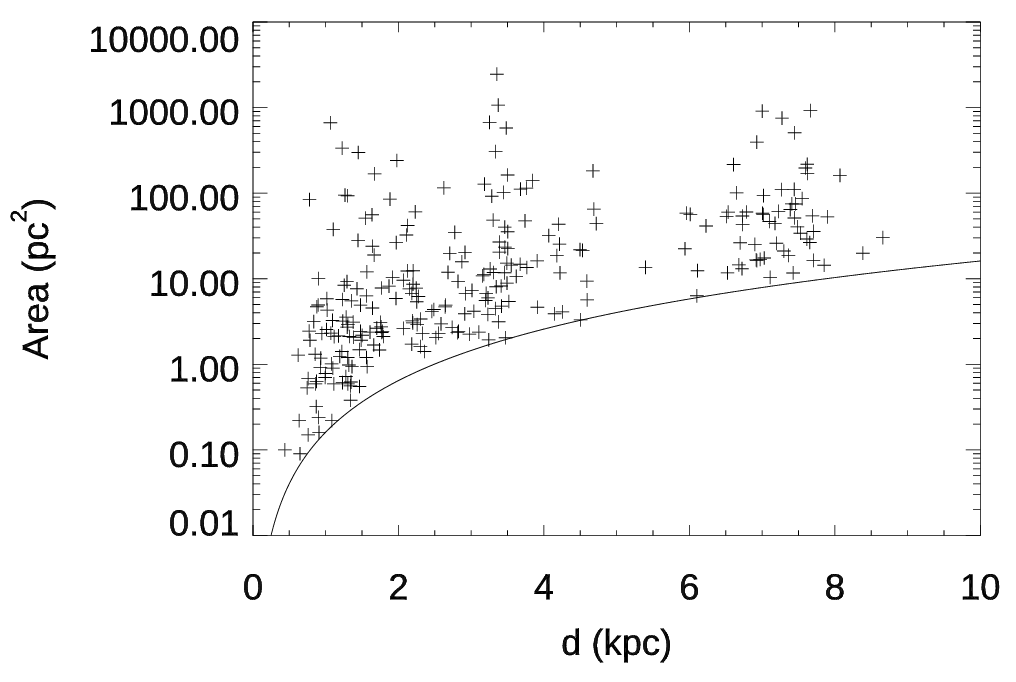}
    \caption{Area of the MCs identified in our survey {\it vs.} their distance. The continuum line defines the limit in the cloud area of two times the beam area; this limit was used as a selection criterion. The possible presence of clouds below this line cannot be established by our survey.
    }
   \label{fig:area-d}
   \end{figure}

We performed the identification of the MCs using the \comain\, line, since it is the brightest line, and the spectral cubes from the filter bank with the largest velocity coverage, namely the FB250. As a first step we masked the FB250 spectral cubes using the dilate mask approach \citep{rosolowsky2006}. In practice, we selected pixels in which the signal is above 5$\times$rms in two consecutive velocity channels and further extended the mask to include all adjacent pixels in which the signal is above 2$\times$rms. We then produced the dendrogram tree of the masked cube without any further filtering of pixels based on signal-to-noise ratio (S/N). This ensures the inclusion of the low-level borders of bright peaks and maximises the connections between structures composed of high-intensity peaks connected by lower intensity levels, while excluding all the pixels with signal below 2$\times$rms and those that are not connected to high-intensity peaks. In practice, we selected structures with S/N$>$5 in at least one spatial pixel and extended their borders up to pixels with S/N=2. To produce the dendrogram tree of the masked cube, we set the minimum difference between two peaks for these to be considered as separate leaves (min\_delta) to 7$\times$rms and the minimum number of pixels of the spectral cube needed for a leaf to be independent to  min\_npix=3$\pi$HPBW$^2$/(4ln2). We finally applied the spectral clustering using the volume of the isosurface in the ({\it l-b-v}) space as a criterion. Since we were also interested in clouds that may simply have little substructure within them, we also included the single leaves of the dendrogram which had been excluded by the clustering algorithm. To be conservative, we selected only single leaves with an angular area larger than two times the beam area, that is, the spatially well-resolved structures. This introduces a minimum physical area of the structures of our catalogue that depends on their distance (see Fig. \ref{fig:area-d}), ranging from 0.16 pc$^2$ at 1 kpc to  10.31 pc$^2$ at 8 kpc.

\begin{figure*}
   \centering
    \includegraphics[width=19.cm]{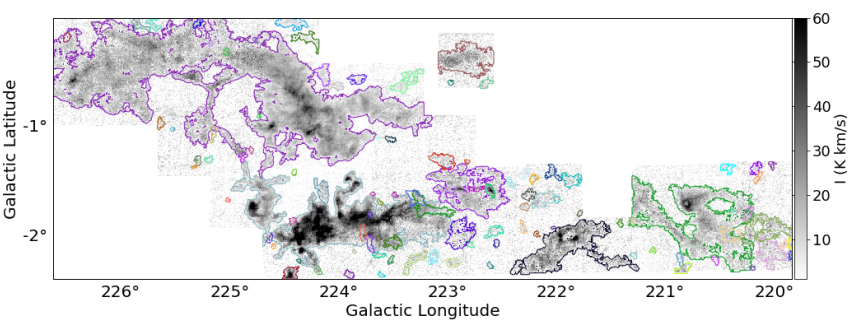}
    \includegraphics[width=19.cm]{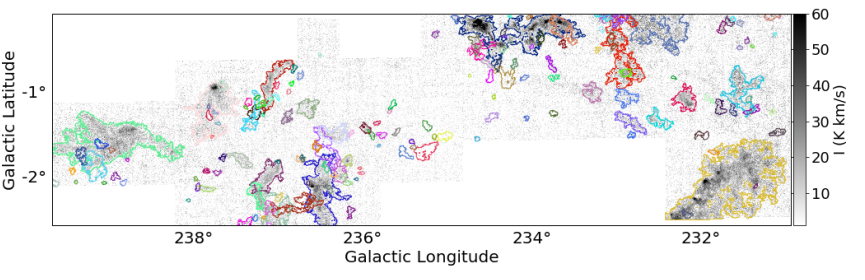}
    \caption{The border of the MCs of the FQS catalogue is drawn over the integrated intensity map of \comain.}
   \label{fig:clouds}
   \end{figure*}

In total, we identified 263 MCs in our \comain\, data cube. A view of the position and contours of the identified structures can be seen in Fig. \ref{fig:clouds}. For each cloud we measured the following properties, using the definition in \citet{rosolowsky2006}: centroid position in Galactic coordinates, semi-major and semi-minor axes, estimated as the rms of the intensity-weighted second moments along the direction of maximum cloud extent and its perpendicular direction, respectively, position angle, intensity-weighted velocity, and relative velocity dispersion ($\sigma_{v}$). These properties are listed in Table \ref{table:MC} for a subsample of MCs. The complete catalogue can be found in the online material. In the catalogue we also give the number of leaves, identified by the dendrogram, that compose the cloud; 235 molecular clouds are composed of only one leaf, indicating that they are relatively homogeneous structures.
  
\begin{sidewaystable*}
\caption{Properties of the molecular clouds identified in the \comain\, data cubes. Only a few objects are listed as example here, the complete catalogue is available at CDS.}             
\label{table:MC}      
\centering

\begin{tabular}{r c c c c c c c c c c c}   
\hline\hline       
index & Name & $l$ & $b$ & $ma$ & $mi$ & PA & $\varv_{\rm lsr}$ & $\sigma_{v}$ & $I_{\rm CO}$ & $L_{\rm CO}$ & $d$ \\
            & & \degr& \degr& \arcsec &\arcsec & \degr & \kms & \kms& K\kms & K\kms pc$^{2}$ & kpc \\
\hline                    
1 &  FQS-MC219.856-2.190 &  219.8564  &  -2.1897  &  74.24  &  22.85 &  84.01 &  28.49  &  0.80  &     10.712  &    13.97   &  2.332 \\
3 &  FQS-MC219.921-2.098 &  219.9211  &  -2.0981  &  75.33  &  42.80 &  52.40 &  31.76  &  1.00  &      8.437  &    26.20   &  2.649 \\
4 &  FQS-MC220.030-1.363 &  220.0299  &  -1.3631  &  74.14  &  45.43 & 113.61 &  17.23  &  0.70  &      5.851  &     3.71   &  1.319 \\
7 &  FQS-MC220.158-1.482 &  220.1583  &  -1.4822  & 119.40  &  31.85 &  55.38 &  70.09  &  1.01  &     22.190  &   229.31   &  7.896 \\

\\
\hline           
\end{tabular}
\vspace{0.5cm}
\begin{tabular}{c c c c c c c c}
\hline\hline       
 $R$ & $A$ & $M$ & $\Sigma$ & $\alpha_{\rm vir}$ &  n$_{\rm l}$ & flag1 & flag2 \\
 pc & pc$^2$ & \msun & \msun pc$^{-2}$ \\
0.889  &  2.30 &      0.00 &   0.00 &     0.00 &     1  &    1  &    2 \\
1.393  &  4.91 &     18.37 &   3.74 &    88.34 &     1  &    1  &    0 \\
0.709  &  0.98 &     44.49 &  45.27 &     9.06 &     1  &    0  &    2 \\
4.509  & 52.83 &   2888.87 &  54.68 &     1.86 &     1  &    0  &    0 \\ 
\hline                    
\end{tabular}

\tablefoot{Columns are as follow. Column 1; progressive index. Column 2; name defined as ``FQS-MC'' followed by the Galactic coordinates of the cloud's centroid. Columns 3 and 4: Galactic longitude and latitude of the cloud's centroid. Columns 5 and 6: intensity-weighted semi-major and semi-minor axes,  $ma$ and $mi$, respectively. Column 7: position angle with respect to the cube x-axis. Column 8: mean velocity. Column 9: velocity dispersion. Column 10: \comain\, integrated intensity across the area of the cloud. Column 11: total luminosity from the \comain\, line. Column 12: kinematic distance from the Sun. Column 13: equivalent radius, assuming circular geometry. Column 14: area. Column 15: total mass derived from \her\, H$_2$ column density. Column 16: average surface density. Column 17: virial parameter. Column 18: number of dendrogram leaves. Column 19: flag1; 0 indicates that the cloud is fully mapped in the \her\, H$_2$ column density map, 1 indicates that the area of the cloud as derived from the \comain\, data is only partially covered, or not covered at all, by the \her\, H$_2$ column density map; for those clouds the derived mass and surface density are lower limits and the virial parameter is an upper limit. Column 20: flag2; 0 means that the cloud is fully mapped in \comain, 2 means that the area of the cloud as derived from the \comain\, touches the border of the map; those clouds could extend outside the mapped area, therefore the measured parameters are linked to an uncertainty that depends on how much CO emission was missed.
}
\end{sidewaystable*}

\subsection{Cloud physical parameters}
\label{sect:phys_par}

From the central velocity of the \comain\, line we derived the kinematic distance of the emitting cloud by applying a Galaxy-rotation model. We used the IDL routine of the CPROPS package \citep{rosolowsky2006} for the distance estimate. This assumes a flat Galactic rotation curve, which is a good approximation for the outer Galaxy, with a solar galactocentric distance of $R_0$ = 8.34 kpc and a rotation velocity of $\Theta_0$ = 240 \kms\, \citep{reid2014}. Since we are looking towards the outer Galaxy, our distance estimate is not affected by the near or far ambiguity.

From the geometrical mean of the semi-axis ($\sigma_r$) we derived the radius of the equivalent spherical cloud: $R = 1.91\sigma_r$ \citep{rosolowsky2006}.
The mass of the clouds was calculated from 
\begin{equation}
\label{eq:mass}
 M = \mu_{\rm H_2} m_{\rm H} \int{N_{\rm H_2} {\rm d}A}
,\end{equation}
where $\mu_{\rm H_2}$ = 2.8 is the molecular weight for the hydrogen molecule that takes into account the presence of helium, $m_{\rm H}$ is the mass of the hydrogen atom, $A$ is the area of the cloud, and $N_{\rm H_2}$ is the molecular hydrogen column density. We have three different possibilities to derive $N_{\rm H_2}$ and therefore the mass of the clouds. The classical method to derive the H$_2$ column density from the \comain\, total integrated intensity is to assume a constant conversion factor between the two:
\begin{equation}
\label{eq:nh2}
N_{\rm H_2} = X_{\rm CO} I_{\rm CO}
.\end{equation}
However, the most recent CO observations of MCs with improved spatial resolution of the order of tens of arcseconds and therefore able to resolve the cloud internal structure, at least for the nearest clouds, have shown that assuming a constant $X_{\rm CO}$ is too simplistic an approach. Indeed, the $X_{\rm CO}$ factor, measured over tens of square degrees in the Galactic plane, shows pixel by pixel variation of more than two orders of magnitude (e.g. \citealt{barnes2015,schuller2017}) due to the combined effect of the change of excitation temperature, optical depth, and CO abundance, in regions with very different physical conditions, even within the same molecular cloud. We therefore preferred to derive the cloud mass from the H$_2$ column density estimated by \her\, by using the far-infrared emission of the cold dust and assuming a gas-to-dust mass ratio. In Sect. \ref{x_fact} we  further discuss the reasons for this decision. The \her\, H$_2$ column density map (Schisano et al., submitted) was derived by fitting the dust emission spectral energy distribution in four far-infrared wavelengths from 160 \um\, to 500 \um, under the assumption of a dust opacity law of $\kappa_{\lambda}=\kappa_{\rm 300} ~ (\lambda/300~\mu m)^{-\beta}$ with $\kappa_{\rm 300}$=0.1 cm$^2$g$^{-1}$ (already accounting for a gas-to-dust ratio of 100) and a grain emissivity parameter $\beta$=2 \citep{hildebrand83}. Being derived from photometric data, the \her\, H$_2$ column density map cannot disentangle clouds at different distances that overlap along the line of sight. We resolved this by distributing the total H$_2$ column density of the \her\, map among all the velocity components identified in the same spatial pixels of the \comain\, cube, when present, proportionally to their \comain\, intensity. This is a correct assumption in the low-density regions of the molecular clouds where the \comain\, line can be optically thin. On the other hand, in the denser regions of the clouds, the \comain\, line is expected to be optically thick and the former assumption introduces an uncertainty in the mass estimate. We stress that since the spatial pixels with a multiple velocity component represent a small percentage of the total survey coverage, 60\% of the MCs of our catalogue are not affected by this problem at all. For the rest of the sample we evaluated this uncertainty by performing the following experiment. We derived an upper limit for the  mass of the clouds using the same \her\, H$_2$ column density map, but in the spatial pixels where multiple velocity components are present, instead of distributing the H$_2$ column density of the \her\, map among the components, we attributed all the H$_2$ column density of that pixel to each of the clouds at the different velocity. We then compared the cloud mass with the upper limit and found that only 11\% of the MCs of our catalogue have a mass that could be underestimated by more than a factor of 0.5, and that  never is the mass underestimated by more than a factor of two. The third possibility to estimate H$_2$ column density is to calculate the CO column density and multiply it by the CO chemical abundance with respect to H$_2$. Having observed the same transition of two isotopologues of carbon monoxide, we can derive an estimate of the CO column density that takes into account the effect of the line opacity. However, this exercise is deferred to a forthcoming paper. 

We used the mass derived from the \her\, H$_2$ column density map to calculate the average mass surface density ($\Sigma$) of the clouds as the ratio between the mass and the area. We then computed the virial parameter, $\alpha_{\rm vir} = \frac{5 \sigma_{\rm v}^2 R}{M G}$, where $\sigma_{\rm v}$ is the gas velocity dispersion, $R$ the spherical radius, $M$ the cloud mass and $G$ the gravitational constant \citep{mckee1992}. 
We derived the velocity dispersion of the gas from the measured full width at half maximum (FWHM) of the \comain\, line through the relation $\sigma_{\rm v}  = FWHM / \sqrt{8 \ln(2)}$. We stress that we accurately measured  the FWHM since the \comain\, line is spectrally resolved in our data. Moreover, since we are looking toward the outer Galaxy where cloud crowding is not an issue, the \comain\, observed lines are well fitted with a single Gaussian profile and our measurement of the gas velocity dispersion of the single MCs is not affected by blending of emission of possible multiple clouds along the same line of sight. It is worth noting that the used relation between FWHM and velocity dispersion is valid for optically thin Gaussian emission line. For an optically thick line, the measured FWHM overestimates the velocity dispersion by a factor that increases with the opacity. In particular, the line broadening due to the opacity becomes larger than 50\% of the measured line broadening for $\tau > $ 10 \citep{hacar2016}. Since the \comain\, is typically optically thick in the densest part of MCs, our estimate of the gas velocity dispersion and, in turn, that of the virial parameter should be considered as an upper limit. 

Distance and distance-dependent parameters of the clouds, namely radius in pc, area in pc$^2$, mass, average surface density, and virial parameter are reported in Table \ref{table:MC}.
We note that some clouds identified in the \comain\, map are only partially mapped or are not mapped at all in the \her\, H$_2$ column density map (see Fig. \ref{fig:coverage}). The mass and the surface density derived for these clouds are lower limits and the virial parameter is an upper limit; these clouds are flagged with flag1 = 1 in Table \ref{table:MC}. Other clouds extend up to the border of the \comain\, map and therefore might not be  completely mapped by our observations. For these clouds, identified with flag2 = 2 in Table \ref{table:MC}, the measured parameters suffer by an uncertainty that depends on how much CO emission was missed.

\subsection{Global properties of the catalogue}
\label{sect:catalog_prop}

  \begin{figure*}
   \centering
    \includegraphics[width=0.9\textwidth]{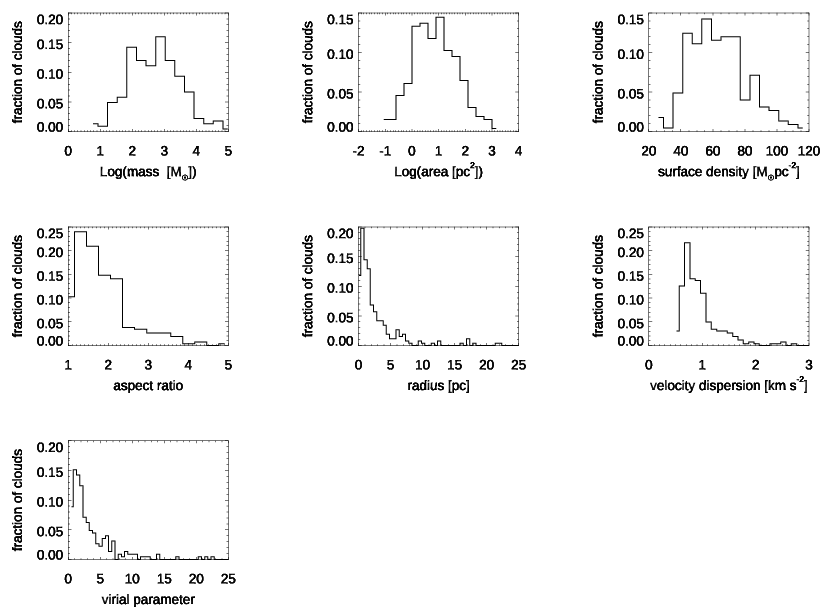}
    \caption{Histograms of the distributions of masses (top-left), area (top-centre), average mass surface densities (top-right), aspect ratios (middle-left), equivalent spherical radius (middle-centre),velocity dispersion (middle-right),  and virial parameters (bottom-left) of the FQS catalogue of molecular clouds.}
   \label{fig:histograms}
   \end{figure*}

%
\begin{table}
\caption{Statistical physical properties of the MCs of the FQS catalogue.}             
\label{tab:statistic}      
\centering                          
\begin{tabular}{l c c}        
\hline\hline                 
    & median & mode \\    
\hline                        
mass (\msun)     & 759 & 760 \\
area (pc$^2$)    & 9.6 & 11.3 \\
surface density (\msun pc$^{-2}$) & 65 & 56 \\
aspect ratio     & 1.8 & 1.3 \\
equivalent spherical radius (pc) & 1.8 & 0.6 \\
velocity dispersion (\kms) & 0.9 & 0.7 \\
virial parameter & 2.51 & 1.06 \\
\hline                                   
\end{tabular}
\end{table}

In the portion of the Milky Way plane covered by our survey, catalogues of MCs were previously produced by \citet{miville2017} and \citet{rice2016} using the survey of \citet{dame2001}.
Thanks to the improved sensitivity and spatial resolution of the FQS survey, we were able to detect many more structures with respect to these two previous catalogues. In particular, in the same surveyed portion of the Galactic plane we identified 263  MCs while \citet{miville2017} found 67 structures and \citet{rice2016} found none. Although limited to about 20\degr\, in Galactic longitude, the FQS MC catalogue represents a good statistical sample of MCs in the outer Galaxy, containing structures distributed in all the three main outer spiral arms up to a distance of $\sim$ 8.6 kpc from the Sun.

The distributions of several physical parameters of the MCs of our catalogue are shown in Fig. \ref{fig:histograms} and their statistical properties are summarised in Table \ref{tab:statistic}. The distributions of area and mass of our sample span over about four orders of magnitude, and when expressed in logarithmic scale they are relatively symmetric with respect to their median value of 9.6 pc$^2$ for the area and 759 \msun\, for the mass. Of course our sample is incomplete at the lowest areas and masses because of the impossibility to reveal small clouds at large distances. In particular, from Fig. \ref{fig:area-d} it is clear that in our catalogue we are able to recover structures with area above $\sim$ 10 pc$^2$ over the entire distance range of 8 kpc. Despite the large range covered by both mass and area, their ratio, that is the average surface density, is less variable, spanning a narrower range from 20 to 120 \msun pc$^{-2}$, with a relatively symmetric distribution with respect to its median value of 65 \msun pc$^{-2}$. This indicates a relatively narrow correlation between the mass and physical dimensions of the clouds. Fitting the area--mass relation with a linear relation in the log-log plane, from the 10 pc$^2$ area limit we can estimate a corresponding limit of the cloud mass of 625 \msun. For both mass and area, these two limits correspond to the turning points of the respective distributions, giving an indication of the completeness limits of the FQS catalogue.

The distributions of aspect ratio, radius, velocity dispersion, and virial parameter are narrow and well peaked, with a little tail toward the highest values. Their modes are: 0.6 pc for the radius, 1.3 for the aspect ratio, 0.7 \kms\, for the velocity dispersion, and 1 for the virial parameter. 

Comparing the properties of our sample with other catalogues extracted from the same tracer such as the \citet{miville2017} and \citet{rice2016} catalogues, we find that our structures are typically smaller and less massive. For example, for the subsample of the 67 clouds of the \citet{miville2017} catalogue with centroid inside the FQS surveyed area, the median value of the radius is 26 pc while in our sample the median value is 1.8 pc and only 13 clouds have radii larger than 10 pc. This is a direct consequence of the improved spatial resolution of our survey that is able to identify also parsec- and subparsec-size clouds that are otherwise blended or undetected in the \citet{dame2001} data, acquired with a telescope beam of 8\farcm4, which is about nine times larger than our beam. This indicates that the new generation of CO surveys, at subarcminute spatial resolution, can detect not only the classical large molecular clouds, that is, molecular structures with radii above 10 pc, but also parsec- and subparsec-sized clouds up to a distance of 2--3 kpc, which means structures that were not detectable in the first generation of CO surveys. This is also confirmed by other MCs catalogues based on subarcminute-spatial-resolution data, such as the GRS \citep{roman-duval2010} and the SEDIGISM \citep{schuller2017} catalogues. Both these latter catalogues, even if covering portions of the Galactic plane inside the solar circle and using another tracer, the \coisot\, line, find a significant number of molecular clouds with radii below a few parsecs. 

One might wonder whether or not the relatively low values of the measured radius of the clouds in our catalogue could be due to an artificial segmentation of larger structures induced by our cloud-identification method. We stress that we tuned the parameters which regulate the cloud identification algorithm in order to maximise the level of clustering of adjacent structures. We also did an a posteriori visual check of the results and we verified that the identified clouds were well-defined structures in our data-set, clearly separated both spatially and spectrally from possible adjacent structures. In a few regions however, it is also possible that some small clouds that appear to be independent structures in our data set, but are spatially close to other structures with similar velocity, could actually be part of a much larger complex not properly traced in its fainter parts by the measured \comain\, line, for sensitivity reasons. This could be the case for the few-tenths-of-a-parsec-sized structures of the catalogue that could represent the high-density clumps of a more extended and less dense cloud. In Sect. \ref{sect:small_clouds} we further discuss this topic.

As an additional check of the reliability of the small MCs, we verified whether or not these clouds are associated to an enhancement of the H$_2$ column density. In particular, we used the following procedure. For all the MCs with  a circular radius below 1 pc and fully mapped in the \her\, map, we selected a region of 2 pixels in width along the border of the area of the cloud as defined in the CO data. We then calculated the median value of the H$_2$ column density in this border and considered this median value as a sort of  H$_2$ column density background level for the cloud. Finally, we calculated the percentage of the area of the cloud where the H$_2$ column density is higher than the background  H$_2$ column density. Values of this percentage above the 50\% indicate that the CO cloud is associated to a systematic enhancement of the H$_2$ column density with respect to the median value in its surrounding. We found that only 5 MCs of the selected 51 small clouds have a percentage below 50\% and the median value of the percentage is 78\%. This result reinforces our conclusions that the small MCs are real overdensities against the underlying diffuse emission. 

\section{Subparsec-sized structures}
\label{sect:small_clouds}

\begin{figure}
   \centering
    \includegraphics[width=9cm]{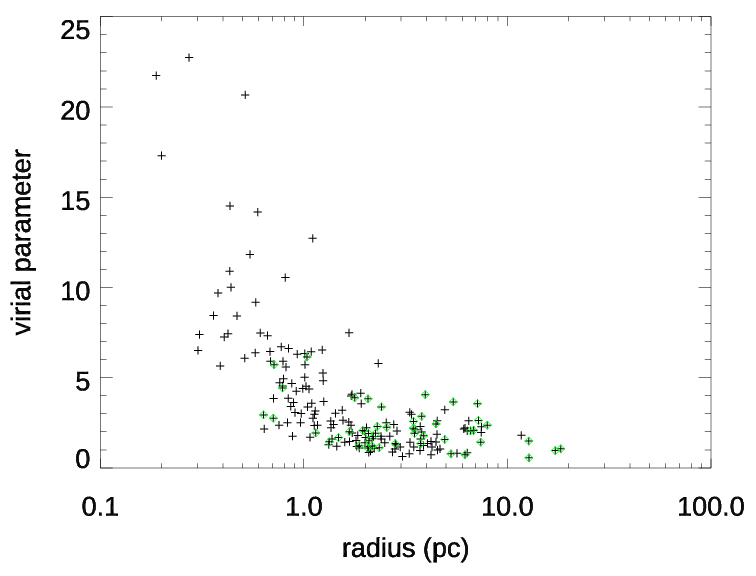}   
    \caption{Radius--virial parameter plot of the MC catalogue. The clouds that are associated with one or more protostar of the \her\, 70 \um\, catalogue are enclosed in a green diamond.}
   \label{fig:aplha-radius}
   \end{figure}

Figure \ref{fig:aplha-radius} shows the virial parameter, estimated from the $^{12}$CO $(1-0)$ line, as a function of cloud radius, coherent with our definition of the cloud radius and mass. We note that since the \comain\, line is typically optically thick in MCs, at least in the denser regions of the clouds, the measured virial parameters are upper limits. In this plot we excluded clouds not fully mapped in \her, for which we did not derive the total mass, and clouds that, touching the border of the \comain\, map, could be more extended. For these latter, both the measured radius and mass must therefore be considered as lower limits. 
We note that the smallest clouds tend to have the largest virial parameter. There is an anti-correlation between these two quantities, with a Pearson's coefficient of -0.7. For a Bonnor-Ebert sphere in isothermal equilibrium and in absence of magnetic field, $\alpha_{\rm vir}\gtrsim2$ is the critical value above which the cloud is gravitationally unbound and may expand and dissolve \citep[e.g.][]{kauffmann2013}, unless confined by external pressure. We found that all the clouds with $R<$ 1 pc have a virial parameter above the critical value of two and that the largest values of the virial parameter ($\alpha_{\rm vir}>5$) are associated only with clouds with $R <$ 2 pc. This seems to indicate that around the radius of 1--2 pc there could be a change in the properties of the clouds. Structures with sizes above 1--2 pc are typical MCs that may be self-gravitating while subparsec-sized structures cannot be kept in by their self-gravity and would be transient or confined by external pressure. 

To further explore the nature of these small structures, we investigated the presence of star formation activity inside them by looking for positional association with \her\,-detected protostars, namely very young protostellar objects still embedded in their parental envelopes and therefore detectable in the catalogue of \her\, 70 \um\, point sources. We found that none of the clouds with $\alpha_{\rm vir} \gtrsim$ 6 and $R \lesssim 2$ pc are associated with a \her\, 70 \um\, point source. We deduced that the sample of small clouds with high virial parameters is likely made up of transient structures not able to activate the star formation process.
  
 \begin{figure}
   \centering
    \includegraphics[width=9cm]{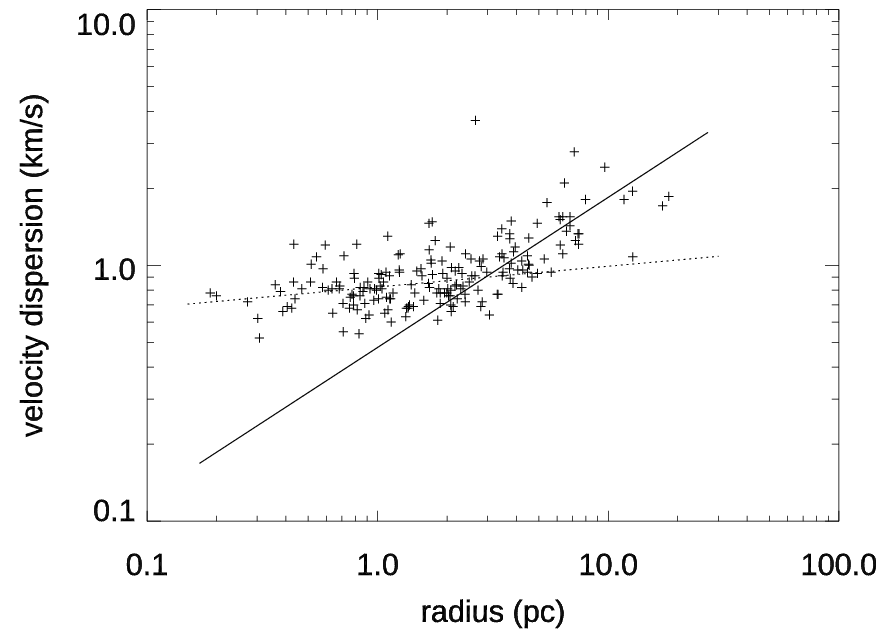}   
    \caption{Radius--velocity dispersion plot of the MC catalogue. The lines represent the best fit of the relation $\sigma_{v} = a R^\beta$. For MCs with $R<$ 2 pc (dotted line) the best-fit coefficients are: $a$ = -0.08 and $\beta$ = 0.08.  For MCs with $R\geq$ 2 pc (solid line) the best-fit coefficients are: $a$ = -0.32 and $\beta =$ 0.59. }
   \label{fig:vel_disp-radius}
   \end{figure}

Another possible indication of the different nature of the smallest clouds with respect to the largest ones is shown by their behaviour in the velocity dispersion versus radius relation. This relation, known as the first Larson relation \citep{larson1981}, is expected to be of the form $\sigma_{v} \propto R^\beta$. The observed non-thermal motions in MCs are always supersonic (thermal motions for gas at T=10 K is $\simeq0.1$ km s$^{-1}$), and the expected value of $\beta$, if the origin of these motions is purely supersonic turbulence, the so-called Burger turbulence, is $\beta=0.5$ \citep{mckee2007}. Various values of $\beta$ in MCs have been derived by several authors, ranging from $\beta \sim$ 0.38 \citep{larson1981} to $\beta =0.6\pm$0.3 \citep{miville2017}. Restricting the evaluation of the relation to the interstellar clouds of the outer Galaxy, previous estimates give $\beta =0.47\pm$0.08 \citep{sodroski1991}, $\beta =0.53\pm$0.03 \citep{brand1995}, $\beta =0.45\pm$0.04 \citep{may1997}, and $\beta =0.53\pm$0.06 for MCs in the third quadrant of the Galaxy \citep{rice2016}, all in agreement with the a Burger-like turbulence. However, the nature of the non-thermal motions in MCs is still debated \citep{krumholz2018}, and it is not yet clear whether they originate solely from turbulence or from large-scale gravitational collapse \citep{ballesteros-paredes2011,traficante2018a,traficante2018b,merello2019} . We plot the $\sigma_{v}$ -- $R$ relation for clouds of the FQS catalogue in Fig. \ref{fig:vel_disp-radius}. Two different regimes appear for clouds with $R\geq$ 2 pc and clouds with $R<2$ pc. For MCs with $R\geq$ 2 pc the velocity dispersion increases with radius with an exponent $\beta$ = 0.59, similar to what is expected from pure supersonic turbulence, and to that previously measured. On the other hand, for MCs with $R<$ 2 pc the relation is almost flat, with $\beta$ = 0.08. The non-thermal motions in the smallest clouds seem to be independent of  cloud size. These condensations, likely to be only transient structures, are probably not formed by the turbulent ISM, which would lead to a Burger-like spectrum. It may be possible that these structures are the result of a large-scale effect, such as spiral-arm density-wave shock that injects the same amount of kinetic energy to all these objects. This large-scale effect would dominate over the local turbulence or the self-gravity within the smallest clouds, becoming increasingly irrelevant in the larger clouds. 

As already noted in Sect. \ref{sect:catalog_prop}, it is also possible that some of the really small structures, namely those with radii of only a few tenths of a parsec, are part of a much bigger complex not properly traced by the \comain\, line. If this is the case, the identified structures would be the densest clumps of a larger cloud. We note that in a survey of massive clumps in the inner Galaxy \citet{traficante2018b} found that the Larson relations valid for MCs are systematically violated. In particular, for these massive clumps, with radii of $0.1\leq R\leq1$ pc, comparable with the size of the small structures we are discussing here, \citet{traficante2018b} found $\beta=0.09\pm0.04$, similar to the value we found for clouds with $R\leq2$ pc. 
A more extended analysis of the dynamics of these clouds in relation to their environment and a comparison of the kinematics of these objects obtained from the $^{12}$CO $(1-0)$ and $^{13}$CO $(1-0)$ lines will be the topic of a forthcoming paper.

\section{Position in the Galaxy}
\label{sect:pos_gal}

  \begin{figure}
  \centering
    \includegraphics[width=8cm]{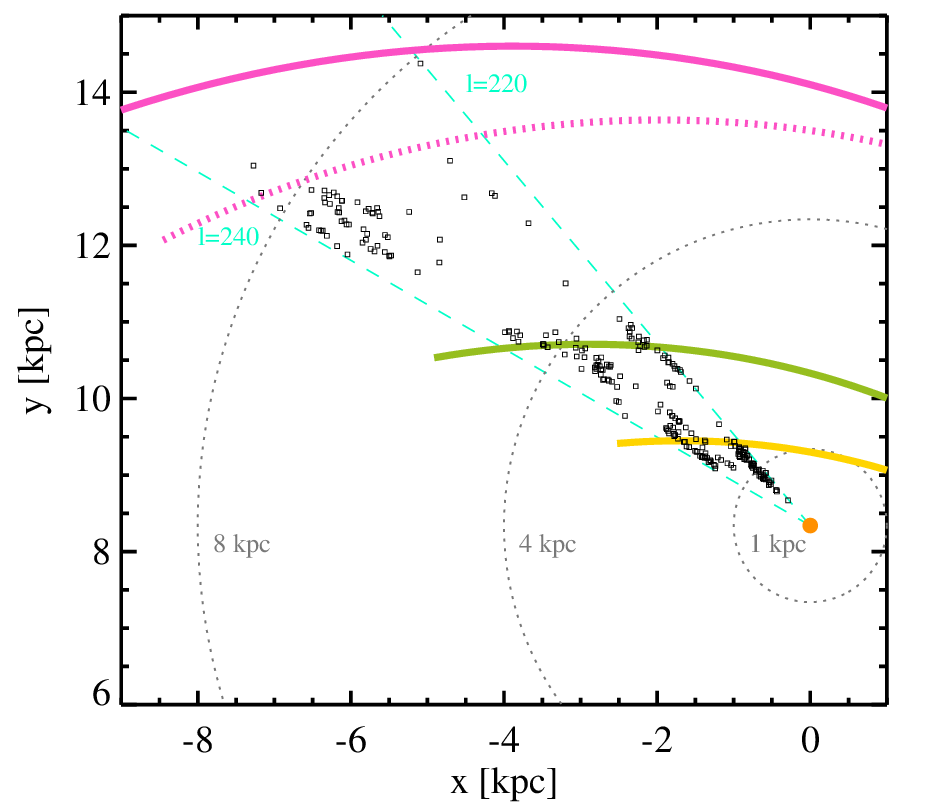}
    \caption{Face-on view of the portion of the Milky Way mapped in the FQS survey where the origin of the (x,y) coordinates is the Galactic centre. The orange point at (0,8.34) kpc marks the position of the Sun \citep{reid2014}. Circles at 1, 4, and 8 kpc from the Sun are indicated by dashed lines. The position of the MCs of the FQS catalogue are represented by squares. The modelled position of three spiral arms is traced by yellow, green, and magenta lines for the Local, Perseus, and Outer arms, respectively. Model parameters for the spiral arms are derived from the fitting of positions of water masers derived from parallax measurements as part of the BeSSeL survey (\citealt{xu2013} for the Local arm, \citealt{choi2014} for the Perseus arm, and \citealt{hachisuka2015} for the Outer arm). An alternative model of the Outer arm \citep{hou2014} is also shown as a magenta dashed line.}
   \label{fig:gal_pos}
   \end{figure}

We used the position and distance of the MCs in the FQS catalogue to describe the structure of the Milky Way in this poorly known portion of the third quadrant. In Fig. \ref{fig:gal_pos} we show the distribution of the clouds projected onto the Milky Way plane. At any Galactic longitude the clouds are grouped in three well-separated structures that trace the position of the spiral arms. The dispersion of the positions in each structure is given not only by the  width of the arms but also by the uncertainty of the distance determination. Indeed, kinematic distances are affected by significant uncertainties due not only to the assumed Galactic rotation model but also to the unknown possible peculiar motions of the clouds themselves. In Fig. \ref{fig:gal_pos} we compare the positions of the clouds derived from FQS data with the expected locations of the spiral arms, adopting a log-periodic spiral form of the Galaxy \citep{reid2014}. The pitch angle and the distance at a reference position of the three arms are taken from \citet{xu2013} for the Local arm, \citet{choi2014} for the Perseus arm, and \citet{hachisuka2015} for the Outer arm. These authors used the positions of water masers derived from parallax measurements as part of the Bar and Spiral Structure Legacy (BeSSeL) survey to fit the parameters of the  spiral arms. We found good agreement between the position of the FQS clouds and the modelled position of the Local and the Perseus arms, while the modelled Outer arm lies further than the observed clouds. We note that while the parameters of the Local and Perseus arms are derived by fitting about 20 water masers located in both the second and third quadrants, covering also the Galactic longitude of the FQS survey, the parameters of the Outer arm are derived from only 5 water masers, 4 of which are located in the second quadrant. Therefore it is not surprising that they might not be fully appropriate to describe the location of the Outer arm in the longitude range of the FQS survey of $220\degr < l < 240\degr$. In fact, an alternative model of the Outer arm from \citet{hou2014}, also shown in Fig. \ref{fig:gal_pos}, seems in better agreement with the FQS results.

\section{ $X_{\rm CO}$ factor}
\label{x_fact}

  \begin{figure}
   \centering
    \includegraphics[width=8cm]{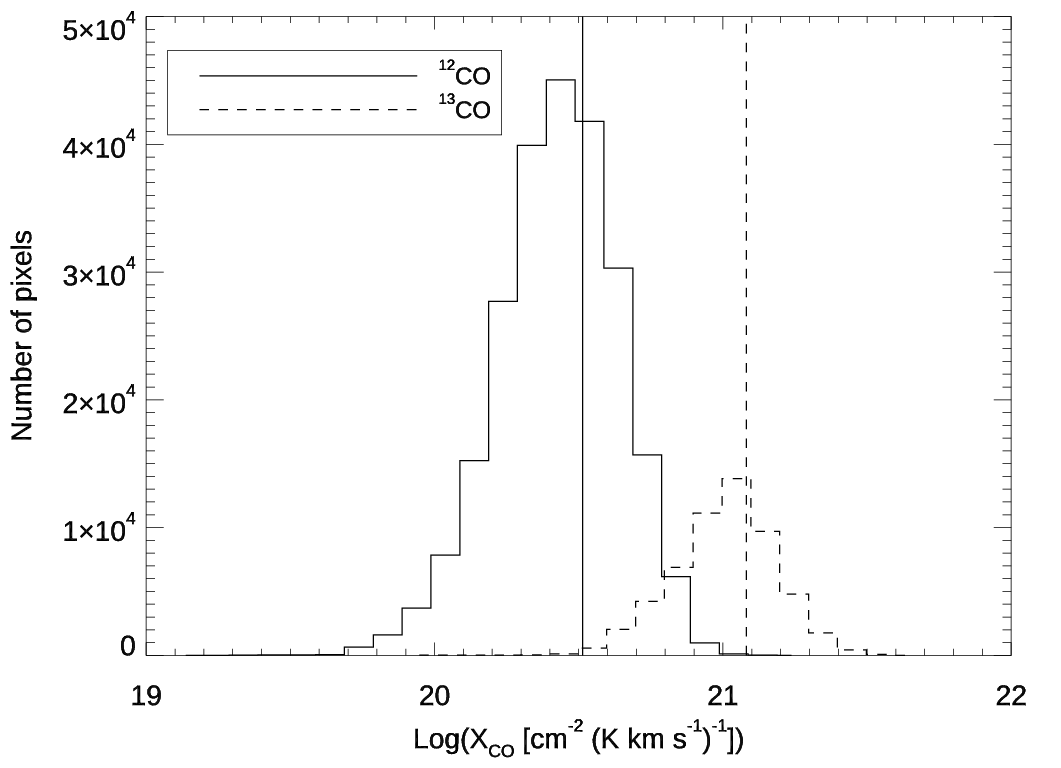}   
    \caption{Histogram of the $X_{\rm CO}$ conversion factor derived from the ratio of the \her\, H$_2$ column density and the line intensity of \comain\, (continuum line) and \coisot\, (dotted line), pixel by pixel. The vertical lines indicate the mean values of 3.3$\times$10$^{20}$\cmdue(K \kms)$^{-1}$ and 1.2$\times$10$^{21}$\cmdue(K \kms)$^{-1}$.}
   \label{fig:xco}
   \end{figure}

We used the ratio between the \her\, H$_2$ column density and the integrated intensity of the CO lines to calculate the $X_{\rm CO}$ conversion factor (Eq. \ref{eq:nh2}) for both \comain\, and \coisot. The histogram of $X_{\rm CO}$ calculated in each spatial pixel of our maps is shown in Fig. \ref{fig:xco}. Although our survey covers a limited portion of the outer Galactic plane of about 20\degr$\times$2.5\degr, we found that the conversion factor is far from being constant, spanning about one order of magnitude from pixel to pixel, indicating that the emitting regions are in very different physical states. This result is in agreement with what is found in other new-generation CO surveys (e.g. \citealt{barnes2015,schuller2017}), where the sub-arcmin resolution allows to resolve the internal structure of the ISM up to distances of a few kiloparsecs, and sample regions with very different values of excitation temperature and optical depth of the \comain\, transition as well as possible variations of the CO abundance. Since a constant conversion factor is not a good approximation for our data, as already mentioned, we preferred to estimate the mass of the MCs of our catalogue from the \her\, data. Indeed, \her\, data have a spatial resolution similar to that of our FQS data and do not suffer from the saturation problems of the \comain\, line that is expected to be optically thick in large parts of the MCs.

On the other hand, a mean value of $X_{\rm CO}$ can still be useful when dealing with low-spatial-resolution data or unresolved structures, and when only CO data are available. We therefore calculated the mean value of the $X_{\rm CO}$ conversion factor for the \comain\, line and we found $X_{\rm CO}$ = (3.3$\pm$1.4)$\times$10$^{20}$ \cmdue(K \kms)$^{-1}$, slightly higher than but still compatible, within the statistical error, with the value of 2$\times$10$^{20}$ \cmdue(K \kms)$^{-1}$ recommended by \citet{bolatto2013} for the Milky Way in their review paper. Interestingly, the $X_{\rm CO}$ derived from FQS data is very similar to that derived in the ThrUMMS survey, a new generation CO (1--0) survey with an effective spatial resolution of 72\arcsec, which finds a median value of 3.8$\times$10$^{20}$ \cmdue(K \kms)$^{-1}$ \citep{barnes2015}. 
For \coisot, we find a mean value of the conversion factor of (1.2$\pm$0.4)$\times$10$^{21}$ \cmdue(K \kms)$^{-1}$, consistent with the value (1.08$\pm$0.19)$\times$10$^{21}$ \cmdue(K \kms)$^{-1}$ found in the science demonstration field of the SEDIGISM survey \citep{schuller2017} at $l$=340.0\degr--341.5\degr.
It is worth noting that, although our method to derive the constant conversion factor is different from that used in the ThrUMMS and SEDIGISM surveys, we find consistent values. Indeed, we derive the H$_2$ column density from  \her\, data that is from the far-infrared emission of the cold dust and assuming a gas-to-dust mass ratio, while both \citet{barnes2015} and \citet{schuller2017} derived the H$_2$ column density from only CO data, measuring the line opacity and excitation temperature from several lines and isotopologues and assuming a CO chemical abundance with respect H$_2$. Moreover, FQS observed only molecular gas outside the solar circle, that is at a galactocentric distance of $R_{\rm gal} \gtrsim$ 8.3 kpc, while both ThrUMMS and SEDIGISM surveys mapped regions mainly inside the solar circle. The similarity of all these estimates of $X_{\rm CO}$, as well the agreement with the estimate of (3.3$\pm$1.7)$\times$10$^{20}$ \cmdue(K \kms)$^{-1}$ based only on \comain\, data in a sample of MCs of the far outer Galaxy (15.7 $ < R_{\rm gal} < $ 20.2 kpc; \citealt{brand1995}) suggests that $X_{\rm CO}$ is invariant, or at least not strongly variable, with galactocentric distance.

\section{Conclusions}

We present the FQS project, a survey in the \comain\, and \coisot\, lines of the Milky Way plane in the range 220\degr$<l<$ 240\degr\, and -2\fdg5$<b<$0\degr, carried out with the 12m ARO antenna.
The median value of rms of the single tiles ranges from 0.79 K to 1.31 K for \comain\, and from 0.33 K to 0.58 K for \coisot\  at the instrumental spectral resolution of 0.65 \kms. We built a mosaic of all the spectra in spectral cubes with spatial pixels of 17\farcs3 and velocity channels of 1 \kms. The median value of the rms in these final spectral cubes is 0.53 K and 0.22 K for \comain\, and \coisot, respectively. 

One of the main goals of the FQS project is to study the large-scale distribution of the ISM in the mapped portion of the outer Galaxy. For this purpose, we produced a catalogue of MCs and studied their global properties. The identification of the MCs and the measurement of their physical parameters were carried out using the \comain\, spectral cubes. We identified 263 MCs, many more than previously found in the same area from the \citet{dame2001} data. The clouds are grouped in three main structures corresponding to the Local, Perseus, and Outer arms of the Galaxy, up to a distance of $\sim$8.6 kpc from the Sun.  This is the first self-consistent statistical catalogue of molecular clouds of the outer Galaxy obtained with a subarcminute spatial resolution. Such a resolution allows us to detect not only the classical giant molecular clouds but also the small clouds and to resolve the cloud structure at the subparsec scale up to a distance of a few kiloparsec.  Indeed, the median value of the spherical equivalent radius of the detected structures is 1.8 pc. The mass of the detected clouds ranges from a few tens to $\sim$10 $^5$ \msun, while the distribution of their average surface density spans a relatively narrow range around the median value of 65 \msun pc$^{-2}$, indicative of a tight correlation between the cloud mass and the size.

The correlation of both the velocity dispersion and the virial parameters with the equivalent spherical radius suggests that there could be a change in the properties of the identified structures at a radius of 1-2 pc. Molecular structures with sizes above a few parsecs are typical MCs that may be self-gravitating, with non-thermal motions dominated by supersonic turbulence. Structures smaller than $\sim$ 1-2~pc are not gravitationally bound and could dissolve unless confined by external pressure. In addition, the velocity dispersion of these clouds does not appear to depend on their size, which suggests an origin in large-scale motions such as density wave shocks in spiral arms.
The smallest structures with radii of only a few tenths of a parsec could represent the high-density clumps of more extended clouds, the most tenuous part of which would be undetected due to lack of sensitivity.

We used the ratio between the H$_2$ column density derived from \her\, data and the integrated intensity of the FQS observed CO lines to calculate the $X_{\rm CO}$ conversion factor for both \comain\, and \coisot. We found that the conversion factor spans about one order of magnitude from pixel to pixel as a consequence of the capacity of our data to resolve the internal structure of the MCs composed of regions in very different physical states. The mean values of $X_{\rm CO}$ are (3.3$\pm$1.4)$\times$10$^{20}$ \cmdue(K \kms)$^{-1}$ and (1.2$\pm$0.4)$\times$10$^{21}$ \cmdue(K \kms)$^{-1}$ for \comain\, and \coisot,\, respectively, which are very similar to the values obtained in other new-generation CO surveys of comparable spatial resolution and in agreement, within the uncertainties, with older estimates.

\begin{acknowledgements}

FQS members thank the ARO 12-m telescope operators: Michael Begam, Kevin Bays, Robert Thompson, Clayton Kyle for their professional assistance and pleasant company during the long observational campaign.
MM acknowledges support from the grant 2017/23708-0, S\~ ao Paulo Research Foundation (FAPESP).
\end{acknowledgements}

\end{document}